\documentclass[12pt]{article}
\usepackage{amsmath, amssymb}
\usepackage{amsfonts}
\usepackage{graphicx,psfrag,epsf}
\usepackage{enumerate}
\usepackage{natbib, xcolor}
\usepackage{url} 
\usepackage[noend]{algpseudocode}
\usepackage{algorithm}
\usepackage{booktabs}
\usepackage{multirow}
\usepackage{verbatim}
\usepackage{array}
\usepackage{caption}
\usepackage{setspace}
\usepackage{authblk}
\usepackage{rotating}
\usepackage{subcaption}
\usepackage[compact]{titlesec}
\usepackage{hyperref}
\usepackage[margin=0.8in]{geometry}
\usepackage{epstopdf}

\newcommand{\blind}{0}

\algnewcommand\And{\textbf{and}}

\DeclareMathOperator*{\argmin}{arg\,min}
\begin{document}

\def\spacingset#1{\renewcommand{\baselinestretch}%
{#1}\small\normalsize} \spacingset{1}


\if0\blind
{

  \title{\bf Evaluating and Testing for Actionable Treatment Effect Heterogeneity}
\author[1]{Mahsa Ashouri}
\author[2]{Nicholas C. Henderson}
\affil[1]{{\small Department of Statistics, Miami University}}
\affil[2]{{\small Department of Biostatistics, University of Michigan, Ann Arbor}}

\date{}
  \maketitle
} \fi

\if1\blind
{
  \bigskip
  \bigskip
  \bigskip
  \begin{center}
    {\LARGE\bf Parameter-Expanded ECME Algorithms for \\[2mm] Logistic and Penalized Logistic Regression}
\end{center}
  \medskip
} \fi

\bigskip
\abstract{
Developing tools for estimating heterogeneous treatment effects (HTE) and individualized treatment effects has been an area 
of active research in recent years. While these tools have proven to be useful in
many contexts, a concern when deploying such methods is
the degree to which incorporating HTE into a prediction
model provides an advantage over predictive methods which do not allow for variation
in treatment effect across individuals. To address this concern, we propose 
a procedure which evaluates the extent to which an HTE model provides a predictive advantage.
Specifically, our procedure targets the gain in predictive performance 
from using a flexible predictive model incorporating
HTE versus an alternative model which is similar to the HTE-utilizing model except that it is
constrained to not allow variation in treatment effect.
By drawing upon recent work in using nested cross-validation techniques for
prediction error inference, we generate confidence intervals for this measure of gain in predictive performance 
which allows one to directly calculate the level at which one is confident 
of a substantial HTE-modeling gain in prediction -- a quantity which we refer to as the h-value.
Our procedure is generic and can be directly used to assess the benefit of modeling HTE for
any method that incorporates treatment effect variation.}

\noindent%
{\it Keywords:} interaction, model comparison, precision medicine, resampling
\vfill

\newpage
\spacingset{1.35} 



\section{Introduction}

Many statistical tools to investigate heterogeneous treatment effects (HTE), to estimate individualized treatment effects,
and to support precision medicine aims more broadly have emerged in recent years. 
Despite these advances, concerns persist about the relatively weak variation in treatment effects and the often insufficient sample sizes to accurately capture this variability. Even when HTE is present in the population of interest, the limited extent of variation and small sample sizes can complicate efforts to create models that significantly improve upon models that do not allow for HTE,
and when considering a new HTE model, it can often be unclear what the predictive value of incorporating
HTE will be. When deploying a new model for patient outcomes in practice, prediction performance is driven
both by the prognostic piece of the model and the part of the model which utilizes HTE,
and determining the added utility of the HTE portion of the model can be a challenge. 
To better evaluate the practical utility of modeling treatment effect variation
in a given HTE analysis, we introduce a procedure that specifically aims to provide an overall assessment 
of the evidence that allowing for treatment variation when modeling patient outcomes offers any predictive benefit.

Statistical tests for detecting the presence of HTE have been proposed in many different approaches to the 
analysis of HTE. In more traditional regression modeling approaches to the analysis of HTE,
variation in treatment effect is modeled through the inclusion
of treatment-covariate interaction terms, and hence, a natural overall test for the presence of HTE is
a likelihood ratio test that the regression coefficients of all treatment-covariate
interaction terms are equal to zero \cite{kovalchik2013}. An alternative approach to HTE analysis is
based on ``risk modeling'' \cite{kent2007, kent2018} where it is assumed that much of HTE can be explained
by variation in a single prognostic, baseline risk score, and in this framework, one can test for HTE by 
testing whether or not a single treatment-risk score interaction parameter equals zero \cite{follmann1999}.
In many applied contexts, subgroup analyses have been widely used as a tool to assess the consistency of a treatment effect
across important patient sub-populations, and reporting tests of treatment effect differences in subgroups 
defined by one patient characteristic at a time is a common way to assess whether or not treatment effects
vary across pre-specified subgroups \cite{alosh2015}.
In the context of subgroup analysis, tests for the presence of ``qualitative interactions''
where the sign of the treatment effect differs from that of the overall treatment 
effect \cite{gail1985} have also been proposed as a way of detecting subgroups that 
could be harmed by the treatment.

Another relevant part of the literature involves nonparametric tests of HTE which rely less on making
particular regression modeling assumptions or on pre-specifying appropriate subgroups. Examples of such nonparametric approaches include the tests described in \cite{crump2008, chang2015, dai2022}. The method outlined in \cite{crump2008} focuses on testing whether or
not the conditional average treatment effect (CATE) function -- the expected difference in response to two 
treatments as a function of baseline covariates -- is constant, and the authors use nonparametric regression techniques
to develop Wald-type tests for this null hypothesis. In \cite{chang2015}, a nonparametric, kernel-based
approach is used to develop a test for the null hypothesis that the CATE function is nonpositive for all covariate values,
and in \cite{dai2022}, a U-statistic-based test is proposed to test the null hypothesis that there is no variation
in the conditional average treatment effect is constant across many strata.

While nonparametric tests for HTE and related procedures offer a useful framework
for detecting the presence of HTE, they do not directly evaluate the ability of a
particular model to harness existing HTE for improved prediction of patient outcomes.
It can often be the case that the magnitude and form of the underlying variation in treatment
effect and the available sample size will severely limit
the ability of many HTE models to usefully harness such variation -- even when HTE is
known to be present in the population of interest.
This motivates our goal of creating an evaluation measure which
does not simply detect the presence of HTE, but rather,
aims to detect the extent to which incorporating HTE within a particular prediction
model leads to improvements in performance when deploying this model 
in a new context.
Specifically, our goal is to create a procedure that compares the out-of-sample performance of an HTE-incorporating
model with the out-of-sample performance of a similar model that is restricted to not
have treatment effect variation. We will consider the difference in
the out-of-sample performance between these two models to be the main predictive
estimand \cite{van2020} of interest, and our objective is to perform statistical inference 
on this quantity. This predictive estimand will depend on the original 
HTE-incorporating model under consideration, and to ensure broad applicability, our procedure
will be completely model agnostic and have the ability to be applied to any approach
incorporating variation in treatment effect.
Flexible machine learning methods have more recently emerged as popular tools
in the analysis of HTE; for example, a recent review \cite{inoue2024} on the use of machine 
learning methods in analyzing HTE in clinical trials
found that causal forests \cite{wager2018, dandl2024, lu2018}, Bayesian additive regression trees \cite{hahn2020, hill2011, zeldow2019}, XGBoost \cite{chen2016, oikonomou2022}, penalized regression, and SuperLearner \cite{van2007super} were among the most popular of such methods. While such methods can estimate treatment
effects at an individual level and generate measures of HTE-related variable importance,
the inferential tools to assess their predictive value relative to machine learning 
procedures which have no HTE are lacking, and it is this gap in assessment measures
that we are aiming to fill.

To perform inference on our predictive estimand of interest,
we propose a two-stage approach. First, for a given HTE-incorporating method
chosen by the researcher, we describe a novel, treatment-shifted outcomes procedure that 
fits the closest non-HTE counterpart of the original model chosen by the researcher.
Second, given that one can generate fitted values from 
the full HTE-incorporating method and the restricted, non-HTE version
of that method, one can directly compute differences in
squared prediction error made from these two models,
and hence, one can use cross-validation-type techniques
to evaluate the out-of-sample difference in predictive performance.
Building on this, we show how one can use nested cross-validation
techniques \cite{varma2006, bates2024} to construct a confidence interval for the predictive difference of interest.
Constructing a confidence interval for this predictive
difference enables one to better assess the evidence for whether or not this predictive
difference is less than zero or less than a threshold of practical significance.
Using nested cross-validation rather than assessing performance on a separate holdout ``validation''
dataset allows the utilization of all the data in the construction
of the confidence interval, which can provide valuable improvements in 
the power to detect differences in predictive performance
in common settings where sample sizes are small relative to the challenges
of building an effective HTE-incorporating model.

This paper is organized as follows. In Section \ref{sec:Method}, we outline our proposed method by first 
describing a general procedure for constructing a restricted conditional mean function estimator
that is restricted in the sense that it satisfies the constraint of no
treatment effect variability across covariate values. Next, in Section \ref{sec:Method}, we describe
our main target for inference, which is the expected difference in the predictive performance
of the restricted and unrestricted estimators of the conditional mean function, and we describe
how to utilize specialized nested cross-validation techniques to construct confidence intervals for this quantity. Section \ref{sec:ModifiedOM} considers alternative, modified outcome methods and describes
how our approach can be adapted to provide inference for the relative predictive performance of
such methods. Section \ref{sec:Sim} includes three simulation studies, involving linear and non-linear outcome models, that evaluate the performance of our proposed approach. Finally, in Section \ref{sec:App}, we demonstrate the performance and use of our method 
through the analysis of outcomes from the Infant Health and Development Program (IHDP) study.

\section{Unrestricted and Zero-HTE-Restricted Outcome Models} \label{sec:Method}
\subsection{Data Structure and Notation}
We consider the setting where one has individual-level data from $n$ participants in a
randomized study with two treatment arms. 
For the $i^{th}$ individual in the study, we let $(Y_{i}(0), Y_{i}(1))$ denote the 
potential outcomes for the outcome of interest. Here, $Y_{i}(0)$ denotes 
the outcome of individual $i$ if individual $i$ was assigned to the control arm of the study, and
$Y_{i}(0)$ denotes the outcome of individual $i$ if individual $i$ was assigned to the active treatment arm.
The treatment assignment indicator variable for individual $i$ is denoted by $A_{i} \in \{0, 1\}$ with $A_{i} = 1$ denoting membership 
in the active treatment arm of the study and $A_{i} = 0$
denoting membership in the control arm of the study. 
The observed outcome for individual $i$ is denoted by $Y_{i}$, and
these outcomes are connected to treatment assignment $A_{i}$ and the random vector $(Y_{i}(0), Y_{i}(1))$ of 
potential outcomes by the following 
\begin{equation}
Y_{i} = A_{i}Y_{i}(1) + (1 - A_{i})Y_{i}(0). \nonumber 
\end{equation}
In addition to observing $Y_{i}$ and $A_{i}$, we observe the $p \times 1$ vector $\mathbf{x}_{i}$ which
contains all baseline covariate information for individual $i$. In total,
we denote the available data by $\mathcal{D} = \{ (Y_{i}, A_{i}, \mathbf{x}_{i}); i=1,\ldots,n \}$.

We assume the main interest is to utilize the dataset $\mathcal{D}$ to build a model for predicting participant
outcomes $Y_{i}$ from treatment and covariate information $(A_{i}, \mathbf{x}_{i})$, and in particular,
we assume a key focus of the model building process is to incorporate treatment-covariate interactions. 
An underlying assumption of incorporating treatment-covariate interactions is
that heterogeneous treatment effects (HTE) play an important role in explaining the variability 
in participant outcomes. 

We let $f(A, \mathbf{x})$ denote the mean function of outcomes $Y_{i}$ conditional 
on treatment assignment $A_{i} = A$ and covariate vector $\mathbf{x}_{i} = \mathbf{x}$.
Namely,
\begin{align}
f(A, \mathbf{x}) &= E\{ Y_{i} \mid A_{i} = A, \mathbf{x}_{i} = \mathbf{x} \} \nonumber \\
&= AE\{Y_{i}(1) \mid \mathbf{x}_{i} = \mathbf{x} \} + (1 - A) E\{ Y_{i}(0) \mid \mathbf{x}_{i} = \mathbf{x} \},
\label{eq:cond_mean_function}
\end{align}
where the second equality follows from the fact that $(Y_{i}(1), Y_{i}(0))$ and treatment assignment $A_{i}$
are assumed to be independent. For example, with continuous outcomes $Y_{i}$, one might assume that the
following outcome model holds
\begin{equation}
Y_{i} = f(A_{i}, \mathbf{x}_{i}) + \varepsilon_{i}, 
\label{eq:continuous_outcomes}
\end{equation}
where $\varepsilon_{i}$ is a mean-zero residual term that is independent of $(Y_{i}, A_{i}, \mathbf{x}_{i})$.
Outcome model (\ref{eq:continuous_outcomes}) could arise, for example, if potential outcomes 
are generated from the following models $Y_{i}(0) = f(0, \mathbf{x}_{i}) + u_{i0}$ and 
$Y_{i}(1) = f(1, \mathbf{x}_{i}) + u_{i1}$, in which case $\varepsilon_{i} = A_{i}u_{i1} + (1 - A_{i})u_{i0}$.

The individualized treatment effect $\theta(\mathbf{x}_{i})$ (ITE) function \cite{rekkas2020} -- often referred to 
as the conditional average treatment (CATE) function \cite{kunzel2019} -- for an individual with baseline covariate 
vector equal to $\mathbf{x}$ is defined as the difference in expected response under the active treatment arm versus the 
expected response under the control arm. The ITE function can be expressed in 
terms of the conditional mean function (\ref{eq:cond_mean_function}) as follows
\begin{eqnarray}
\theta( \mathbf{x} ) 
&=& E\{ Y_{i} \mid A_{i} = 1, \mathbf{x}_{i} = \mathbf{x} \} - E\{ Y_{i} \mid A_{i}=0, \mathbf{x}_{i} = \mathbf{x} \}  \nonumber \\
&=& f(1, \mathbf{x}) - f(0, \mathbf{x}). \nonumber 
\end{eqnarray}
We will say that HTE is present if the ITE function $\theta( \mathbf{x} )$ does not depend on $\mathbf{x}$, i.e., when $\theta(\mathbf{x})$ equals a constant term for all $\mathbf{x}$. Hence, outcome models which aim to incorporate HTE
are either explicitly or implicitly specifying an ITE function which varies with $\mathbf{x}$ while
models which do not allow for HTE assume that $\theta(\mathbf{x})$ equals some constant
value for all $\mathbf{x}$.

\subsection{Treatment-shifted Outcomes and the Restricted Baseline Risk Estimator}
The key issue we want to address is the extent to which modeling HTE improves the predictive performance
of an estimated outcome model compared to an estimation method which does not take HTE into account. 
More specifically, we want to compare the predictive performance associated with an unrestricted estimator 
of the conditional mean function $f(A, \mathbf{x})$ versus an alternative modeling strategy which resembles the
unrestricted estimator except for the fact that it does not incorporate HTE into the modeling strategy. 
For the unrestricted estimation approach, we use $\hat{f}(A, \mathbf{x}; \mathcal{D})$
to denote the full-dataset \textit{unrestricted estimate} of $f(A, \mathbf{x})$. This estimator
is unrestricted in the sense that it does not place any restrictions on the ITE function $\theta(\mathbf{x}) = f(1, \mathbf{x}) - f(0, \mathbf{x})$.  We will compare the predictive performance of $\hat{f}(A, \mathbf{x})$
with a competing \textit{restricted estimator} $\hat{g}(A, \mathbf{x}; \mathcal{D})$
that is restricted in the sense that the difference 
$\hat{g}(1, \mathbf{x};\mathcal{D}) - \hat{g}(0, \mathbf{x}; \mathcal{D})$ 
cannot depend on $\mathbf{x}$. The restricted estimator $\hat{g}(A, \mathbf{x}; \mathcal{D})$ can potentially be a very flexible estimator of the mean function $f(A, \mathbf{x})$, but the associated ITE function $\hat{g}(1, \mathbf{x};\mathcal{D}) - \hat{g}(0, \mathbf{x}; \mathcal{D})$ must be constant.
To better formalize this notion, we assume that a restricted estimator must have the following form
\begin{equation}
\hat{g}(A, \mathbf{x}; \mathcal{D}) = \hat{f}_{\tau^{*}}(\mathbf{x}; \mathcal{M}_{\tau^{*}}) + \tau^{*} A. 
\label{eq:restricted_fn_form}
\end{equation}

The form of the restricted estimator in (\ref{eq:restricted_fn_form}) ensures that the ITE function 
is constant because $\hat{f}_{\tau^{*}}(\mathbf{x}; \mathcal{M}_{\tau^{*}})$ does not depend on $A$ 
and hence $\hat{g}(1, \mathbf{x}; \mathcal{D}) - \hat{g}(0, \mathbf{x}; \mathcal{D})$ equals 
the estimated constant treatment effect term $\tau^{*}$ for all $\mathbf{x}$.
In (\ref{eq:restricted_fn_form}), the term $\hat{f}_{\tau^{*}}(\mathbf{x}; \mathcal{M}_{\tau^{*}})$ represents a 
restricted, ``$\tau^{*}$ -adjusted baseline risk estimate'', i.e., a fitted value for an individual in the control arm with covariate
vector $\mathbf{x}$, under an assumption that there is no HTE and that the true treatment effect is $\tau^{*}$.
To both ensure that the estimated function $\hat{f}_{\tau}(\mathbf{x}, \mathcal{M}_{\tau})$ in (\ref{eq:restricted_fn_form}) does not depend on $A$ and can be interpreted as an adjusted baseline risk estimator,
we assume that this estimate is not obtained through ``training'' on the full dataset $\mathcal{D}$, but rather, $\hat{f}_{\tau}(\mathbf{x}, \mathcal{M}_{\tau})$ is constructed on a derived dataset $\mathcal{M}_{\tau}$ which has treatment-shifted responses as the outcomes but does not have treatment assignment as a covariate. More specifically, $\mathcal{M}_{\tau}$ is the dataset comprising shifted outcomes $Y_{i} - \tau A_{i}$ and all covariates $\mathbf{x}_{i}$ except for treatment assignment. That is, $\mathcal{M}_{\tau} = \big\{ (Y_{i} - \tau A_{i}, \mathbf{x}_{i}); i = 1, \ldots, n \big\}$. The reason for training $\hat{f}_{\tau}(\mathbf{x}, \mathcal{M}_{\tau})$ on the shifted-outcomes dataset $\mathcal{M}_{\tau}$ is to use all of the observations to estimate the baseline risk function (under the assumption of constant treatment effect $\tau$) and to force this function to not incorporate 
any treatment assignment information when learning this baseline risk function.

In (\ref{eq:restricted_fn_form}), an estimated constant treatment effect $\tau^{*}$ is used. 
This estimate should represent the best value of the treatment effect when assuming a restricted 
estimate of the form $\hat{f}_{\tau}(\mathbf{x}; \mathcal{M}_{\tau}) + \tau A$ will be used.
To this end, we define the estimate $\tau^{*}$ as the value of $\tau$ which minimizes the sum of squares error 
between the outcomes $Y_{i}$ and the values of the restricted estimates $\hat{g}(\mathbf{x}_{i}, A_{i}; \mathcal{D})$
associated with $\tau$. Namely,
\begin{equation}
\tau^{*} = \argmin_{\tau} \sum_{i=1}^{n}\Big( Y_{i} - \tau A_{i} - \hat{f}_{\tau}(\mathbf{x}_{i}; \mathcal{M}_{\tau})  \Big)^{2}.
\label{eq:reduced_trt_effect}
\end{equation}
Because $\tau^{*}$ represents a treatment effect estimate under the restriction that there can be no HTE, 
we refer to $\tau^{*}$ as the \textrm{restricted treatment effect estimate}.

\begin{figure}[ht]
		\centering
		\includegraphics[width=16cm, height=12cm]{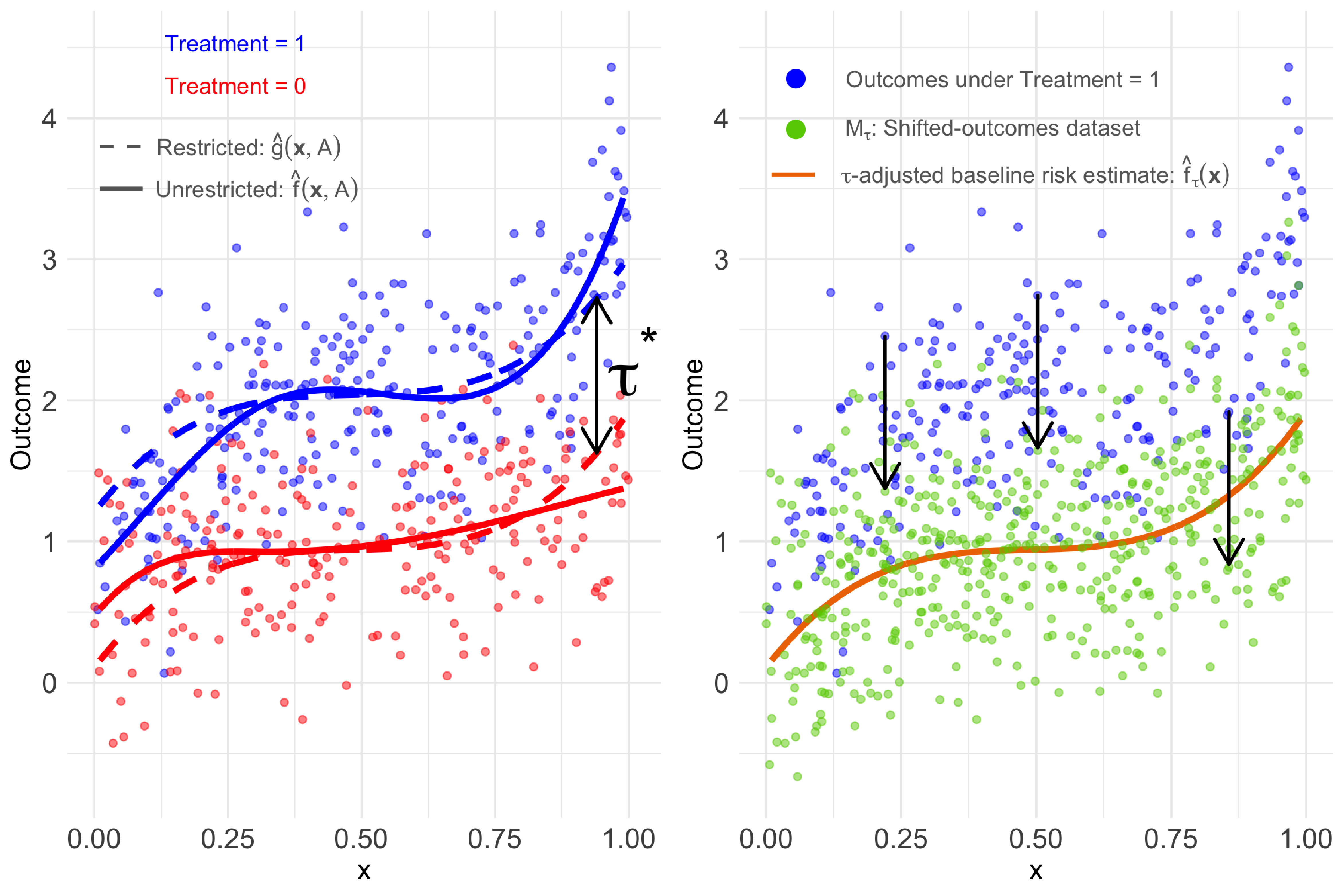}
		\caption{Simulated data illustrating the $\tau$-shifted dataset, computation of the restricted treatment effect estimate $\tau^{*}$, and the restricted estimator $\hat{g}(A, \mathbf{x})$ of the conditional mean function. The downward arrows show how outcome values in the active treatment arm are transformed into outcomes in the shifted-outcomes dataset $\mathcal{M}_{\tau}$.}
		\label{fig:example-plot}
\end{figure}

Figure \ref{fig:example-plot} uses simulated data to provide a conceptual illustration
of the roles of the unrestricted estimator, the restricted estimator, and the shifted-outcomes
dataset $\mathcal{M}_{\tau}$. As shown in the left-hand panel of this figure, the unrestricted estimate $\hat{f}(\mathbf{x}, A;\mathcal{D})$ is, for each value of $A$, a flexible, nonlinear function of the covariate $x$, and 
the estimated ITE function $\hat{f}(1,x) - \hat{f}(0, x)$ varies across the range of $x$. The left-hand panel of Figure \ref{fig:example-plot} shows that the restricted
estimate $\hat{g}(x, A; \mathcal{D})$ is also a flexible, nonlinear function of $x$, but the estimated 
differences $\hat{g}(1, x; \mathcal{D}) - \hat{g}(0, x; \mathcal{D})$ between the treatment arms are constant across the entire range of $x$. The right-hand panel illustrates how the modified outcome dataset is used to construct the $\tau$-adjusted baseline risk estimate $\hat{f}_{\tau}(\mathbf{x}; \mathcal{M}_{\tau})$. The green points in the right-hand panel of Figure \ref{fig:example-plot} represent the $\tau$-shifted outcomes $Y_{i} - \tau A_{i}$ while the blue points represent 
the original outcomes in the active treatment group, and these values are not used when calculating the
$\tau$-adjusted baseline estimate.

\begin{table}
\small
\centering
\begin{tabular}[ht]{ll}
\toprule
Notation&  Definition \\
\midrule
 $\mathcal{D}$ & Full dataset including the treatment assignment variable \\
 $f(A, \mathbf{x})$ & Conditional mean function \\
 $\theta(\mathbf{x})$ & Individualized treatment effect (ITE) function \\
$\mathcal{M}_{\tau}$ & Shifted-outcomes dataset that excludes treatment assignment \\
                     & variable but responses are $Y_{i} - \tau A_{i}$ \\
 $\hat{f}(A, \mathbf{x}; \mathcal{D})$ & Unrestricted estimator of the conditional mean function \\
 $\hat{g}(A, \mathbf{x}; \mathcal{D})$ & Restricted estimator of the conditional mean function \\
$\hat{f}_{\tau}(\mathbf{x}; \mathcal{M}_{\tau})$ & $\tau$-adjusted restricted baseline risk estimator \\
$\tau^{*}$ & Restricted treatment effect estimate \\
$\hat{\theta}(\mathbf{x}; \mathcal{D})$ & Unrestricted estimator of $\theta(\mathbf{x})$ for modified outcome approaches \\
$\tau_{MO}^{*}$ & Restricted estimator of $\theta(\mathbf{x})$ for modified outcome approaches \\
 \bottomrule
\end{tabular}
\caption{Glossary of key terms and notation used.}
\label{tab:summary_terms} 
\end{table}

It is worth noting that the restricted estimator can be directly computed for any procedure that one may choose for estimating the $\tau$-adjusted baseline risk estimators -- including procedures that do not have a direct linear model representation and are viewed as more ``black box'' procedures. In this sense,
the restricted estimator $\hat{g}(A, \mathbf{x})$ can be thought of as a generalization of a linear regression model that allows for a treatment main effect but no treatment-covariate interactions. Indeed, as the following example illustrates, if one is using a direct linear regression approach to estimate the conditional 
mean function $f(A, \mathbf{x})$, then following
our procedure for computing the restricted estimate $\hat{g}(A, \mathbf{x}; \mathcal{D})$ leads to the exact same estimate
of the conditional mean function as fitting a linear 
regression model with only a main effect for treatment.

\bigskip

\noindent \textit{Example: Linear regression with treatment-covariate interaction and a scalar-valued covariate.}  
To better illustrate the procedure described above, we consider an example where one has scalar-valued covariates $x_{1}, \ldots, x_{n}$ and that one is planning to use a direct linear regression
model with a treatment-covariate interaction for the unrestricted estimate of the conditional mean function. 
In this setting, the unrestricted estimator and $\tau$-adjusted baseline risk estimator would take
the form:
\begin{eqnarray}
\hat{f}(A, x; \mathcal{D}) &=& \hat{\beta}_{0} + \hat{\beta}_{1}x + \hat{\beta}_{2}A + \hat{\beta}_{3}Ax \nonumber \\
\hat{f}_{\tau}(x; \mathcal{M}_{\tau}) &=& \hat{\gamma}_{0,\tau} + \hat{\gamma}_{1, \tau}x 
\end{eqnarray}
where $(\hat{\beta}_{0}, \hat{\beta}_{1}, \hat{\beta}_{2}, \hat{\beta}_{3})$ are the least-squares estimates of 
$(\beta_{0}, \beta_{1}, \beta_{2}, \beta_{3})$  
with outcomes $Y_{i}$ and conditional mean function $\beta_{0} + \beta_{1}x_{i} + \beta_{2}A_{i} + \beta_{3}A_{i}x_{i}$ and 
where $(\hat{\gamma}_{0,\tau}, \hat{\gamma}_{1,\tau})$ are the least-squares estimates of $(\gamma_{0,\tau}, \gamma_{1,\tau})$
in the model with outcomes $Y_{i} - \tau A_{i}$ and mean model $\gamma_{0,\tau} + \gamma_{1,\tau} x_{i}$.
Specifically, the values of $\hat{\gamma}_{0,\tau}$ and $\hat{\gamma}_{1,\tau}$ are given by
\begin{equation}
\hat{\gamma}_{0,\tau} = \bar{Y} - \tau\bar{A} - \hat{\gamma}_{1,\tau}\bar{x} \qquad\qquad \hat{\gamma}_{1,\tau} = \frac{\sum_{i=1}^{n}(x_{i} - \bar{x})(Y_{i} - A_{i}\tau) }{\sum_{i=1}^{n} (x_{i} - \bar{x})^{2}}
= C_{xy} - C_{xa}\tau
\end{equation}
where $C_{xy} = \sum_{i=1}^{n} (x_{i} - \bar{x})Y_{i} \big/ \sum_{i=1}^{n} (x_{i} - \bar{x})^{2}$ and 
$C_{xa} = \sum_{i=1}^{n} (x_{i} - \bar{x})A_{i} \big/ \sum_{i=1}^{n} (x_{i} - \bar{x})^{2}$. Using (\ref{eq:reduced_trt_effect}), 
the reduced treatment effect estimate $\tau^{*}$ is obtained by 
\begin{align}
\tau^{*} &= \argmin_{\tau} \sum_{i=1}^{n} \{ Y_{i} - \tau A_{i} - \hat{f}_{\tau}(\mathbf{x}_{i}; \mathcal{M}_{\tau}) \}^{2} \nonumber \\
&= \argmin_{\tau} \sum_{i=1}^{n} \Big\{ (Y_{i} - \bar{Y}) - C_{xy}(x_{i} - \bar{x}) - \tau[(A_{i} - \bar{A}) - C_{xa}(x_{i} - \bar{x})] \Big\}^{2}.
\label{eq:linear_tau_example}
\end{align}
One can verify that the reduced treatment effect estimate $\tau^{*}$ in (\ref{eq:linear_tau_example}) 
is the exact same as the least-squares estimate
of $\tau$ in the outcome model $Y_{i} = \beta_{0} + \beta_{1}x_{i} + \tau A_{i} + \varepsilon_{i}$,
where there is no treatment-covariate interaction term (see Appendix B for a justification of this).

\subsection{Confidence Intervals for Differences in Unrestricted and Restricted Predictive Performance}

We are mainly interested in comparing the out-of-sample performance of the unrestricted and 
restricted estimation procedures. Specifically, if we were to observe a new observation $Y_{n+1}$ with
associated covariate and treatment information $(A_{n+1}, \mathbf{x}_{n+1})$, our interest is in how 
well the estimators $\hat{f}$ or $\hat{g}$ would perform in using $(A_{n+1}, \mathbf{x}_{n+1})$ to predict $Y_{n+1}$. 
To evaluate the relative performance in predicting an out-of-sample outcome $Y_{n+1}$, our aim is 
to conduct inference on the following quantity
\begin{equation}
\theta_{XY} = E\Big\{ \ell\Big( \hat{f}(\mathbf{x}_{n+1}, A_{n+1}; \mathcal{D}), \hat{g}(\mathbf{x}_{n+1}, A_{n+1}; \mathcal{D}), Y_{n+1} \Big) \Big| \mathcal{D} \Big\}. 
\label{eq:theta_equation} 
\end{equation}
In (\ref{eq:theta_equation}), $\ell$ is a loss function that compares the performance of the unrestricted and restricted
estimators in predicting the new outcome $Y_{n+1}$. In other words, $\theta_{XY}$ quantifies the expected relative performance, conditional 
on the observed data $\mathcal{D}$, of the unrestricted and restricted estimators in their ability to generalize
to unobserved data points $(Y_{n+1}, A_{n+1}, \mathbf{x}_{n+1})$ that are drawn from the same distribution that generated observations in $\mathcal{D}$.
For continuous outcomes $Y_{i}$, a natural choice of the comparison loss function $\ell$ is the difference in the squared error between the full and reduced models. This loss function can be expressed as
\begin{multline}
\ell\Big( \hat{f}(\mathbf{x}_{n+1}, A_{n+1}; \mathcal{D}), \hat{g}(\mathbf{x}_{n+1}, A_{n+1}; \mathcal{D}), Y_{n+1} \Big)
= \Big( Y_{n+1} - \hat{f}(\mathbf{x}_{n+1}, A_{n+1}; \mathcal{D}) \Big)^{2} \\
- \Big( Y_{n+1} - \hat{g}(\mathbf{x}_{n+1}, A_{n+1}; \mathcal{D}) \Big)^{2}.
\label{eq:main_loss_function}
\end{multline}
With the above loss function, values of $\ell$ less than zero imply better predictive performance 
of the unrestricted estimator $\hat{f}$ while values of $\ell$ greater than zero imply superior
predictive performance of the restricted estimator $\hat{g}$.

Our main aim is to estimate $\theta_{XY}$ and construct a confidence interval for this predictive estimand. 
A confidence interval for $\theta_{XY}$ can be used to quantify the evidence for whether the restricted or unrestricted
estimator is superior in the predictive sense captured by (\ref{eq:theta_equation}). 
If the upper limit of a confidence interval for $\theta_{XY}$ is less than zero, this would 
provide evidence suggesting a model incorporating HTE has better performance in predicting 
patient outcomes than a model which assumes a constant treatment effect.
To construct the confidence interval for $\theta_{XY}$, we adopt the nested cross-validation scheme 
proposed in \cite{bates2024} where the dataset is repeatedly divided into random ``outer'' training and test sets
and K-fold cross-validation is performed within each of the outer training sets. 
This procedure is meant to address the underestimation of the naive estimator of variance
associated with the common K-fold cross-validation estimate of prediction error which arises from 
inappropriate modeling of the dependence of test-set errors across different folds.

Following the procedure described in \cite{bates2024}, we define our $100\times(1-\alpha)\%$ confidence interval $C_{\alpha}( \mathcal{D} )$ for $\theta_{XY}$ as
\begin{equation}
C_{\alpha}(\mathcal{D}) = \Bigg( \hat{E}^{(NCV)} - \widehat{\textrm{b}}^{(NCV)} - z_{1 - \alpha/2}\sqrt{\widehat{MSE}^{(NCV)}}, \quad
\hat{E}^{(NCV)} - \widehat{\textrm{b}}^{(NCV)} + z_{1 - \alpha/2}\sqrt{\widehat{MSE}^{(NCV)}} \Bigg), 
\label{eq:confint_formula}
\end{equation}
where $z_{1 - \alpha/2}$ denotes the $1 - \alpha/2$ quantile of the standard normal distribution.
In (\ref{eq:confint_formula}), $\hat{E}^{(NCV)}$ is the ``direct'' nested cross-validation estimate
of $\theta_{XY}$. With $R$ ``repetitions'' and $K$ folds, $\hat{E}^{(NCV)}$ is computed by
\begin{equation}
\hat{E}^{(NCV)} = \frac{1}{K(K-1)R}\sum_{r=1}^{R} \sum_{k=1}^{K} \sum_{j=1}^{K-1} \frac{1}{n_{jkr}}\sum_{i \in \mathcal{I}_{jkr}}  L_{ijkr}.\nonumber 
\end{equation}
Above, $n_{jkr}$ is the number of observations in the $j^{th}$ fold $\mathcal{I}_{jkr}$ of the $k^{th}$ ``outer training set'' $\mathcal{A}_{kr}$, and $L_{ijkr}$ is defined as
\begin{equation}
L_{ijkr} =\ell\Big( \hat{f}(\mathbf{x}_{i}, A_{i}; \mathcal{A}_{kr} \backslash \mathcal{I}_{jkr} ), \hat{g}(\mathbf{x}_{i}, A_{i}; \mathcal{A}_{kr} \backslash \mathcal{I}_{jkr}), Y_{i} \Big) \quad \text{ if } i \in \mathcal{I}_{jkr}, \text{ for } j = 1,\ldots, K-1, \nonumber 
\end{equation}
where $\mathcal{I}_{jkr}$ is the $j^{th}$ fold of the $k^{th}$ ``outer training set'' $\mathcal{A}_{kr}$. 
The term $\hat{b}^{(NCV)}$ in (\ref{eq:confint_formula}) is a bias correction described in \cite{bates2024}
that is obtained from the difference in the nested cross-validation estimate $\hat{\theta}^{(NCV)}$ and the usual $K$-fold cross-validation estimate of $\theta_{XY}$. The term $\widehat{MSE}^{(NCV)}$ is a nested cross-validation estimate
of the mean-squared error (MSE) of the K-fold cross-validation estimate of $\theta_{XY}$, and 
the steps to compute the MSE estimate $\widehat{MSE}^{(NCV)}$
are described in more detail in \cite{bates2024}.

The confidence intervals $C_{\alpha}(\mathcal{D})$ naturally lead to an accompanying p-value type quantity that
we refer to as the ``h-value''. If we express (\ref{eq:confint_formula}) as
$C_{\alpha}(\mathcal{D}) = \big( L_{\alpha}(\mathcal{D}), U_{\alpha}(\mathcal{D}) \big)$,
then the h-value is defined as
\begin{eqnarray}
h &=& \inf\Big\{ \alpha \in (0, 1): \textrm{sign}(L_{\alpha}(\mathcal{D})) = \textrm{sign}(U_{\alpha}(\mathcal{D})) \Big\} \nonumber \\
&=& 2\Phi\Bigg(  -\Big| \hat{E}^{(NCV)} - \widehat{\textrm{b}}^{(NCV)} \Big| \Big/ \sqrt{\widehat{MSE}^{(NCV)}}  \Bigg),
\end{eqnarray}
where $\Phi(\cdot)$ is the cumulative distribution function of a standard normal distribution. In other words, 
the h-value represents the smallest value of $\alpha$ at which the confidence interval $C_{\alpha}(\mathcal{D})$
excludes zero. 

The confidence interval in (\ref{eq:confint_formula}) is two-sided, but in the context of
evaluating the benefit of modeling HTE,
one-sided confidence intervals and h-values will often be of more interest. In cases
where the sole focus is on examining whether or not the HTE model has better predictive
performance than the restricted model, one would only want to report the level-$\alpha$ one-sided confidence interval
$\tilde{C}_{\alpha}(\mathcal{D}) = (-\infty, U_{2\alpha}(\mathcal{D}) )$ and conclude that incorporating
HTE in the predictive model can yield superior performance if $U_{2\alpha}( \mathcal{D} ) < 0$.
The one-sided h-value associated with this one-sided confidence interval is
\begin{eqnarray}
\tilde{h}(\mathcal{D}) &=& \inf\Big\{ \alpha \in (0, 1): U_{2\alpha}(\mathcal{D}) < 0 \Big\} \nonumber \\
&=& \Phi\Bigg( -(\hat{E}^{(NCV)} - \widehat{\textrm{b}}^{(NCV)}) \Big/ \sqrt{\widehat{MSE}^{(NCV)}}  \Bigg). \nonumber 
\end{eqnarray}

\section{Modified Outcome Methods}\label{sec:ModifiedOM}

A number of methods for estimating the ITE function $\theta(\mathbf{x})$ do not estimate 
an outcome model for both treatment arms, but only generate an estimate of $\theta(\mathbf{x})$. 
Examples of such methods include causal forests methods \cite{athey2016, wager2018, cui2023}, 
and the ``R-learner'' approach \cite{nie2021}. 
Another family of methods that perform direct estimation of $\theta(\mathbf{x})$ use ``modified outcomes'' to estimate
parameters of a model that does not have treatment assignment as a covariate. The modified outcomes are typically 
constructed so that they have expectation equal to $\theta(\mathbf{x}_{i})$ so that 
a regression model with these modified outcomes as responses and the $\mathbf{x}_{i}$ as
covariates lead to a direct estimate of the ITE function \cite{tian2014, weisberg2015}. 
To be more specific, in the context of a randomized trial, ``modified outcomes'' $W_{i}$ are
often defined as
\begin{equation}
W_{i} = 
\begin{cases}
Y_{i}/P\{ A_{i} = 1 \mid \mathbf{x}_{i} \} & \text{ if } A_{i} = 1 \\
-Y_{i}/P\{ A_{i} = 0 \mid \mathbf{x}_{i} \} & \text{ if } A_{i} = 0.
\end{cases}
\label{eq:modfied_outcomes_definition}
\end{equation}
The modified outcome $W_{i}$ has expectation equal to $\theta(\mathbf{x}_{i})$ which may be
seen by noting that 
\begin{eqnarray}
E\{ W_{i} \mid \mathbf{x}_{i}\} &=& E\{ W_{i} \mid \mathbf{x}_{i}, A_{i} = 1\}P\{A_{i}=1 \mid \mathbf{x}_{i} \}
+ E\{ W_{i} \mid \mathbf{x}_{i}, A_{i} = 0\}P\{A_{i}=0 \mid \mathbf{x}_{i} \} \nonumber \\
&=& E\{ Y_{i} \mid \mathbf{x}_{i}, A_{i} = 1\}
- E\{ Y_{i} \mid \mathbf{x}_{i}, A_{i} = 0\} \nonumber \\
&=& f(1, \mathbf{x}_{i}) - f(0, \mathbf{x}_{i}). \nonumber 
\end{eqnarray}
In our simulations, we only consider modified outcomes of the form (\ref{eq:modfied_outcomes_definition}),
but there are alternative, closely related versions of modified outcomes (\ref{eq:modfied_outcomes_definition})
designed to reduce variance \cite{wu2018} that one could also use. Such modified outcomes still possess
the key property of having expectation equal to $\theta(\mathbf{x}_{i})$.

In the modified outcomes setting, one directly uses the modified outcomes $W_{i}$ and covariates $\mathbf{x}_{i}$ to build a model that estimates $\theta(\mathbf{x}_{i})$. In this setting, an unrestricted estimator
of $\theta(\mathbf{x}_{i})$ is one that places no restrictions on the form of an estimator of $\theta(\mathbf{x})$,
and we denote the unrestricted estimator as $\hat{\theta}(\mathbf{x}; \mathcal{D})$.
Because a restricted estimator of $\theta(\mathbf{x})$ cannot vary with $\mathbf{x}$,
it must be a constant function, and hence,
we define the restricted estimator in this setting to be the constant term $\tau_{MO}^{*}$
which is defined as
\begin{equation}
\tau_{MO}^{*} = \argmin_{\tau} \sum_{i=1}^{n}( W_{i} - \tau)^{2} = \bar{W}. \nonumber 
\end{equation}

To evaluate the performance of an HTE model constructed using a modified outcome method, 
we will compare the predictive performance of the unrestricted ITE function estimator 
$\hat{\theta}(\mathbf{x}; \mathcal{D})$ with that of the constant treatment effect estimate $\tau_{MO}^{*}$.
To do this comparison, we will use a modified version of the loss 
function (\ref{eq:main_loss_function}) that replaces $Y_{n+1}$
with $W_{n+1}$, replaces $\hat{f}$ with $\hat{\theta}$, and replaces $\hat{g}$ with $\tau_{MO}^{*}$.
The modified loss function $\tilde{\ell}$ is defined as
\begin{equation}
\tilde{\ell}\Big( \hat{\theta}(\mathbf{x}_{n+1}; \mathcal{D}), \tau_{MO}^{*}, W_{n+1} \Big)
= \Big( W_{n+1} - \hat{\theta}(\mathbf{x}_{n+1}; \mathcal{D}) \Big)^{2} 
- \Big( W_{n+1} - \tau_{MO}^{*} \Big)^{2},
\label{eq:modified_loss_function}
\end{equation}
and for this modified loss, we define the new predictive estimand $\tilde{\theta}_{XY}$ as the expectation of this modified loss function, conditional on the observed data
\begin{equation}
\tilde{\theta}_{XY} = E\Big\{ \tilde{\ell}\Big( \hat{\theta}(\mathbf{x}_{n+1}; \mathcal{D}), \tau_{MO}^{*}, W_{n+1} \Big) \Big| \mathcal{D} \Big\}, 
\label{eq:tilde_theta}
\end{equation}

One can think of $\tilde{\theta}_{XY}$ as the modified-outcomes counterpart to the predictive performance
measure $\theta_{XY}$ in (\ref{eq:theta_equation}).
The predictive estimand (\ref{eq:tilde_theta}) also has a nice interpretation 
as the difference between $\hat{\theta}(\mathbf{x}; \mathcal{D})$ and $\tau_{MO}^{*}$ in their predictive mean-squared error for the ITE function evaluated at the future covariate value $\mathbf{x}_{n+1}$.
Specifically, one can show using a direct argument (see Appendix A) that 
the predictive estimand $\tilde{\theta}_{XY}$ is equal to 
\begin{equation}
\tilde{\theta}_{XY}
= E\Big\{ \Big( \theta(\mathbf{x}_{n+1}) - \hat{\theta}(\mathbf{x}_{n+1}; \mathcal{D}) \Big)^{2} \Big| \mathcal{D} \Big\}
- E\Big\{ \Big( \theta(\mathbf{x}_{n+1}) - \tau_{MO}^{*} \Big)^{2} \Big| \mathcal{D} \Big\}. 
\label{eq:modified_estimand}
\end{equation}
It is worth noting that one is not limited to use the modified loss (\ref{eq:modified_loss_function}) only for modified outcomes methods,
and can use $\tilde{\ell}$ with unrestricted and restricted estimators of the conditional mean functions. 
Specifically, for unrestricted $\hat{f}(A, \mathbf{x}; \mathcal{D})$ and restricted $\hat{g}(\mathbf{x}, A; \mathcal{D})$
estimators of the conditional mean function, one would use $\hat{\theta}(\mathbf{x}_{n+1}; \mathcal{D}) = \hat{f}(1, \mathbf{x}_{n+1}; \mathcal{D}) - \hat{f}(0, \mathbf{x}_{n+1}; \mathcal{D})$ and replace $\tau_{MO}^{*}$ with $\hat{g}(1, \mathbf{x}_{n+1}; \mathcal{D}) - \hat{g}(0, \mathbf{x}_{n+1}; \mathcal{D})$ in loss (\ref{eq:modified_loss_function}).

\section{Simulation studies}\label{sec:Sim}

In this section, we demonstrate the performance of our proposed methodology using three simulation studies. In the first study, 
we generate outcomes from a model which has a linear regression for the conditional mean function and has normally distributed residuals.
The second simulation study uses a range of more complex nonlinear functions for the conditional mean function
with twelve different combinations of the baseline risk function and ITE function. 
In the third simulation study, we evaluate the use of modified outcome methods on data simulated from the settings of the second simulation study. 
For the first simulation study, we consider sample sizes of $n = 100$ and $n = 500$, and 
in the second simulation study, we consider sample sizes of $n = 500$ and $n = 2,000$.

In these simulation studies, we evaluated coverage and h-values for the predictive estimands $\theta_{XY}$ associated with the 
following four methods: simple linear regression model, $L_{1}$-penalized regression (Lasso) \cite{friedman2010}, $L_{2}$-penalized regression (ridge regression), and boosting \citep{hofner2014model}. Random forest \citep{ho1995random} was also evaluated for several simulation settings, and the results are presented in Appendix C. 

For each setting of the conditional mean function, we generated $500$ separate datasets of size $n$, and
for each dataset and method, we computed the predictive estimands ($\theta_{XY}$ or $\tilde{\theta}_{XY}$), constructed
confidence intervals, and computed h-values. For each generated dataset, we 
evaluated the width of the confidence interval $C_{\alpha}(\mathcal{D})$ and 
recorded whether or not the confidence intervals contained
the predictive estimand $\theta_{XY}$. Coverage results were obtained by computing the proportion
of datasets where $\theta_{XY}$ (or $\tilde{\theta}_{XY}$) was contained in $C_{\alpha}(\mathcal{D})$.
For all methods, five folds were used as the basis for each nested cross-validation repetition,
and $50$ nested cross-validation repetitions were run to construct the confidence intervals.

\subsection{Linear conditional mean function}\label{sim1-linear}

In the first simulation study, we generated outcomes $Y_{i}$ from the following linear model
\begin{equation}
Y_{i} = \beta_{0} + \beta_{1} x_{i1} + \beta_{2} x_{i2} + \beta_{3}A_{i} + \beta_{4} A_{i}x_{i1} + \beta_{5} A_{i}x_{i2} + \varepsilon_{i}, \quad i = 1, \ldots, n, \label{eq:linear_sim_model} 
\end{equation}
where $\varepsilon_{i} \sim \textrm{Normal}(0, 1)$. The treatment assignment variables $A_{i}$
were generated independently as $A_{i} \sim \textrm{Bernoulli}(0.5)$,
and the covariates were generated independently as $x_{ij} \sim \textrm{Normal}(0,1)$.
For the linear model (\ref{eq:linear_sim_model}), we considered two choices of the regression coefficients,
which we label simulation settings $A$ and $B$. These simulation settings used the following parameter values:
\begin{itemize}
\item
\textit{Setting A}: $\beta_0 = 2, \beta_1 = 3, \beta_2 = -1, \beta_3 = 1.5, \beta_4 = 0 , \beta_5 = 0$
\item 
\textit{Setting B}: $\beta_0 = 2, \beta_1 = 3, \beta_2 = -1, \beta_3 = 1.5, \beta_4 = 0.5, \beta_5 = -2$
\end{itemize}
In simulation setting $A$, there is no HTE as the treatment effect is equal to $\beta_{3}$ for all individuals
while, in simulation setting $B$, HTE is present with the ITE function
$\theta(\mathbf{x}_{i}) = \beta_{4}x_{i1} + \beta_{5}x_{i2}$.

Table \ref{tab:simulation1} presents the simulation results for settings A and B with sample sizes of $100$ and $500$. Results are shown 
for unrestricted estimators based on linear regression, Lasso, and boosting. Average values of the data-dependent
predictive estimand $\theta_{XY}$ are reported for each simulation setting and sample size. A positive value of $\theta_{XY}$
indicates that the restricted estimator has better predictive performance than the HTE-incorporating, unrestricted model. 
For setting $B$, the predictive advantage of incorporating HTE methods is clear as the average value
of $\theta_{XY}$ is close to $-1$ for each method, and in setting A where there is no HTE, 
the predictive advantage of modeling HTE is very close to zero for all methods.

For both simulation settings, the coverage of all methods achieved the 95\% nominal level and even had slightly greater 
than $95\%$ coverage in some cases, and in both simulation settings, linear regression and boosting generally have narrower confidence intervals than the Lasso. The larger h-values for all methods in setting A reflect the absence of HTE in this setting with 
the median one-sided h-values ranging from $0.43$ to $0.84$. 
All methods have considerable power to ``detect'' the predictive HTE present in setting B. Even when $n = 100$,
the median one-sided h-values are less than $0.005$ for all methods. 

\begin{table}[ht]
\centering
\begin{tabular}[ht]{lclrccc}
\multicolumn{1}{c}{} & \multicolumn{2}{c}{ } \\
\toprule
n & Simulation & Method & $\theta_{XY}$ & Coverage & CI Width & Median \\
& Setting & &  & Proportion & &  h-value \\
\midrule
100 & $A$ & linear  & 0.022 & 0.986 & 0.270 & 0.637\\
& & Lasso & 0.012  & 0.980 & 0.439 & 0.838\\
& & boosting & 0.017 & 0.992 & 0.185 & 0.657\\
 \midrule
 100 & $B$ & linear  & -1.102 & 0.950 & 1.355 & 0.001 \\
& & Lasso & -1.248 & 0.952 & 1.665 & 0.002 \\
& & boosting & -1.080 & 0.948 & 1.302 &0.001\\
 \midrule
 500 & $A$ & linear & 0.004 & 0.986 & 0.044& 0.609\\
& & Lasso &  0.001 & 0.972 & 0.148 & 0.871\\
& & boosting & 0.006 & 0.980 & 0.032 & 0.432\\
 \midrule
 500 & $B$ & linear  & -1.071 & 0.954 & 0.507 &$<$ 0.001\\
& & Lasso & -1.126 & 0.964 & 0.581 & $<$ 0.001\\
& & boosting & -1.053 & 0.950 & 0.493 & $<$ 0.001\\
 \midrule
\end{tabular}
\caption{Simulation results for the two settings of the linear conditional mean function. Linear regression, Lasso, and boosting methods are compared for sample sizes of $n=100$ and $n=500$. The average value of the predictive estimand $\theta_{XY}$ across simulation replications is reported, and the coverage
and average width of the nested cross-validation based confidence intervals for $\theta_{XY}$ are shown for each method and setting. 
Median one-sided h-values across simulation replications are displayed for each method and simulation setting. 
}\label{tab:simulation1}
\end{table}

\subsection{Non-linear conditional mean functions}\label{sim2}

In our second simulation study, we generate outcomes $Y_{i}$ from the following model
\begin{equation}\label{eq:sim2}
Y_{i} = \mu(\mathbf{x}_{i}) + A_{i}\theta(\mathbf{x}_{i}) + \varepsilon_{i}, \quad i=1, \ldots,n,
\end{equation}
where $\varepsilon_i \sim \textrm{Normal}(0, 1)$ and where the covariate vectors $\mathbf{x}_{i} = (x_{i1}, \ldots, x_{i9})$ have length $9$ and are composed of both continuous and binary covariates. Specifically, we generate
$x_{ij} \sim \textrm{Normal}(0, 1)$ when $j$ is even and generate $x_{ij} \sim \textrm{Bernoulli}(0.5)$ when $j$ is odd.
As in the linear conditional mean function simulation study, the treatment assignment variables $A_{i}$ are generated as $A_{i} \sim \textrm{Bernoulli}(0.5)$.

To simulate outcomes from (\ref{eq:sim2}), we considered three choices of the baseline risk function $\mu(\mathbf{x})$ and four choices of the ITE function $\theta(\mathbf{x})$.
Similar to the simulation design in \cite{powers2018}, we considered the following three settings of the baseline risk function
\begin{equation}\label{eq:mu-sim2}
\begin{aligned}
 \mu_{1}( \mathbf{x}_{i} ) &= x_{i2}x_{i4}x_{i6} + 2x_{i2}x_{i4}(1-x_{i6}) + 3x_{i2}(1-x_{i4})x_{i6} + 4x_{i2}(1-x_{i4})(1-x_{i4})\\
 & + 5(1-x_{i2})x_{i4}x_{i6} + 6(1-x_{i2})x_{i4}(1-x_{i6}) + 7(1-x_{i2})(1-x_{i4})x_{i6} \\
 & + 8(1-x_{i2})(1-x_{i4})(1-x_{i6}),\\ 
 \mu_{2}( \mathbf{x}_{i} ) &= 4I(x_1 > 1) I( x_3>0 ) + 4I(x_5 > 1) I(x_7 > 0) + 2x_8x_9,\\
 \mu_{3}( \mathbf{x}_{i} ) &= \frac{1}{2} ( x_1^2 + x_2 + x_3^2 +x_4 + x_5^2 + x_6 + x_7^2 + x_8 + x_9^2 -11), \\
\end{aligned}
\end{equation}
and we considered the following four options for the ITE function $\theta(\cdot)$
\begin{equation}\label{eq:theta-sim2}
\begin{aligned}
    \theta_1( \mathbf{x}_{i} ) &= 0,\\
    \theta_2( \mathbf{x}_{i} ) &= 1,\\
    \theta_3( \mathbf{x}_{i} ) &= 2 + 0.1/\{1 + \exp(-x_{i2})\},\\
    \theta_4( \mathbf{x}_{i} ) &= \mu_{1}( \mathbf{x}_{i} ).
    \end{aligned}
\end{equation}

With the first choice $\theta_{1}(\mathbf{x})$ of the ITE function, there is both no HTE and an average treatment effect equal to zero. For the second choice $\theta_{2}( \mathbf{x} )$ of the ITE function, there is a positive average treatment effect present, but there is no HTE. With the third setting $\theta_{3}(\mathbf{x})$ of the ITE function, HTE is present, but the variation in treatment effect is rather weak and is only driven by variation in the second covariate $x_{i2}$. HTE is also present in the fourth setting $\theta_{4}(\mathbf{x})$, but the variation in the ITE function depends on a broader collection of covariates than in the third setting of the ITE function. 
Note that there is no HTE present in the first two settings
of the ITE function, and hence, we should expect the predictive difference confidence intervals $C_{\alpha}(\mathcal{D})$ will often contain zero. For the last two settings, we should expect that the confidence intervals $C_{\alpha}(\mathcal{D})$ will often exclude zero, provided that the sample size is sufficiently large and $\theta_{XY}$ is sufficiently far from zero.



Table \ref{tab:sim2} presents the same evaluation measures as Table \ref{tab:simulation1}, and the table 
presents results for all combinations of the baseline risk functions (\ref{eq:mu-sim2}) and 
the ITE functions (\ref{eq:theta-sim2}).  
For settings where the ITE function is set to $\theta_{1}(\mathbf{x})$ or $\theta_{2}(\mathbf{x})$,
the values of $\theta_{XY}$ are mostly positive or close to zero for all methods indicating that,
even with these nonlinear baseline risk functions,
the restricted methods are better or roughly the same as the unrestricted procedures. 
One thing that stands out from Table \ref{tab:sim2} is the considerable variation in $\theta_{XY}$
across methods. Unlike the linear conditional mean function simulation study, 
linear regression and Lasso are misspecified models in 
this context, and hence the degree to which an unrestricted estimator can improve
upon a restricted estimator is not as clear as in the linear setting and will 
depend on how close a linear fit can get to the underlying nonlinear conditional mean function.
A notable example of this is the simulation setting with baseline risk function $\mu_{1}(\mathbf{x})$
and ITE function $\theta_{4}(\mathbf{x})$ where the restricted estimator performance is much better than 
that of the unrestricted estimator. In this setting, the ITE function $\theta_{4}(\mathbf{x})$ 
is quite complicated and simply modeling the ITE function as a constant is, for these sample
sizes, much better than aiming to estimate a linear approximation to the true, nonlinear ITE function.

Regarding the coverage proportions, nested cross-validation with linear regression and boosting have
the most consistent performance. Both of these methods consistently have $95\%$ or greater coverage across 
all combinations of $\mu(\cdot)$ and $\theta(\cdot)$. The Lasso also had generally good coverage performance with coverage
near $95\%$ for most cases. However, in the scenarios that had zero 
treatment effect (i.e., when the ITE function is $\theta_{1}(\mathbf{x})$), Lasso often had less than nominal coverage
with coverage somewhat less than $90\%$ in these scenarios.

In all settings with no HTE (i.e., those with ITE function $\theta_{1}(\mathbf{x})$ and $\theta_{2}(\mathbf{x})$),
the proportion of h-values less than $0.05$ is very small and the median one-sided h-values are typically greater than
$0.5$ and are greater than $0.25$ in all of the $\theta_{1}(\mathbf{x})$ and $\theta_{2}(\mathbf{x})$ settings. 
Thus, in settings with no HTE, none of the methods ``detect'' a predictive 
advantage from incorporating HTE. Indeed, in these simulation settings,
most of the methods are quite conservative in the sense that they do not show evidence for a predictive advantage
of HTE modeling even for many of the settings where HTE is present. 
When $n = 500$, the predictive estimands $\theta_{XY}$ are mostly positive or close to zero for all methods and
settings, indicating that, for this sample size, the restricted estimators typically have at least as good predictive performance as the
unrestricted estimators. Because of this, the median one-sided h-values are quite large for all methods and all simulation settings when $n=500$. It is only when $n=2000$ that we see a few settings where the predictive estimands are clearly negative and
the corresponding h-values become close to zero. There is, however, notable heterogeneity across methods as to which 
choices of $\mu(\mathbf{x})$ or $\theta(\mathbf{x})$ lead to a strong predictive advantage for the unrestricted
estimator for that method. 
For example, with boosting, only the settings with ITE function $\theta_{4}(\mathbf{x})$ 
have a predictive estimand which is negative and far away from zero, and hence, this is the only case
where this procedure yields small one-sided h-values. In contrast, for the linear regression method
with both $\theta_{3}(\mathbf{x})$ and $\theta_{4}(\mathbf{x})$, 
the average of the predictive estimands $\theta_{XY}$ is negative enough to yield small median h-values
except when the baseline risk function is $\mu_{1}(\mathbf{x})$.

\begin{table}[ht]
    \centering
\scriptsize
    \begin{tabular}{cc|rcrc|rcrc|rcrc}
    \multicolumn{7}{c}{} & \multicolumn{2}{c}{n = 500}& \multicolumn{5}{c}{}\\
    \midrule
   \multicolumn{2}{c}{} & \multicolumn{4}{c}{Method = linear}& \multicolumn{4}{c}{Method = Lasso}& \multicolumn{4}{c}{Method = boosting}\\
   \midrule
 $\mu(\cdot)$ & $\theta(\cdot)$ &$\theta_{XY}$ & Coverage & CI & Median$^*$& $\theta_{XY}$ & Coverage & CI & Median$^*$& $\theta_{XY}$ & Coverage & CI & Median$^*$\\
  & & & Prop. & Width & h-value & & Prop. &  Width& h-value& & Prop. &  Width& h-value\\
 \midrule
 $\mu_1$ & $\theta_1$& 2.262&0.992&13.919&0.773& -0.327 &0.962&3.404&0.364& 0.359&0.984&7.283&0.646\\
 $\mu_1$ & $\theta_2$& 2.210&0.996&14.044&0.768& -0.035&0.976&3.864&0.454&0.349&0.990&7.217&0.643\\
 $\mu_1$ & $\theta_3$& 2.253&0.990&14.092&0.766& 1.971&0.958&6.372&0.896&0.409&0.986&7.377&0.644\\
 $\mu_1$ & $\theta_4$& 4.709&0.974&36.289&0.714& 13.182&0.940&24.607&0.988&1.283&0.972&24.508&0.615\\
 $\mu_2$ & $\theta_1$& 0.019&0.986&0.105&0.781&0.002&0.864&0.092&0.554&0.001&0.984&0.033&0.510\\
 $\mu_2$ & $\theta_2$& 0.019&0.992&0.103&0.802& 0.008&0.976&0.159&0.585&0.050&0.978&0.091&0.988\\
 $\mu_2$ & $\theta_3$& -0.024&0.978&0.131&0.261&-0.037&0.974&0.186& 0.223&0.051&0.964&0.124&0.943\\
 $\mu_2$ & $\theta_4$& -0.182&0.984&7.818&0.477&7.169&0.956&12.499&0.981&-0.433&0.982&6.064&0.425\\
 $\mu_3$ & $\theta_1$& 0.019&0.990&0.105&0.785&0.007&0.878&0.102&0.635&0.001&0.978&0.040&0.525\\
 $\mu_3$ & $\theta_2$& 0.019&0.988&0.104&0.784&0.011&0.966&0.155&0.624&0.059&0.968&0.093&0.995\\
 $\mu_3$ & $\theta_3$& -0.025&0.984&0.133&0.250&-0.033&0.986&0.189&0.254&0.067&0.978&0.130&0.981\\
 $\mu_3$ & $\theta_4$& -0.104&0.984&8.415&0.488&8.680&0.948&12.343&0.995&-0.184&0.978&6.337&0.480\\
 \midrule
 \multicolumn{7}{c}{} & \multicolumn{2}{c}{n = 2000}& \multicolumn{5}{c}{}\\
    \midrule
 $\mu_{1}$ & $\theta_{1}$ & 0.236 & 0.992 & 1.418 & 0.771 & -0.142 & 0.846 & 1.539 & 0.275 & 0.037 & 0.992 & 0.713 & 0.642 \\ 
 $\mu_{1}$ & $\theta_{2}$ & 0.224 & 0.986 & 1.371 & 0.797 & -0.022 & 0.852 & 1.958 & 0.520 & 0.057 & 0.996 & 0.732 & 0.685 \\ 
 $\mu_{1}$ & $\theta_{3}$ & 0.190 & 0.990 & 1.428 & 0.738 & 0.788 & 0.912 & 3.756 & 0.851 & 0.028 & 0.988 & 0.771 & 0.615 \\ 
 $\mu_{1}$ & $\theta_{4}$ & -0.694 & 0.984 & 4.342 & 0.280 & 4.021 & 0.856 & 11.364 & 0.928 & -0.958 & 0.980 & 3.431 & 0.154 \\ 
 $\mu_{2}$ & $\theta_{1}$ & 0.002 & 0.992 & 0.009 & 0.799 & 0.000 & 0.784 & 0.013 & 0.463 & 0.000 & 0.980 & 0.003 & 0.358 \\ 
 $\mu_{2}$ & $\theta_{2}$ & 0.002 & 0.986 & 0.009 & 0.805 & 0.000 & 0.962 & 0.023 & 0.538 & 0.042 & 0.966 & 0.022 & 1.000 \\ 
 $\mu_{2}$ & $\theta_{3}$ & -0.041 & 0.970 & 0.027 & $< 0.001$ & -0.043 & 0.978 & 0.047 & $< 0.001$ & 0.037 & 0.970 & 0.031 & 1.000 \\ 
 $\mu_{2}$ & $\theta_{4}$ & -1.139 & 0.970 & 1.412 & 0.001 & 0.841 & 0.882 & 2.745 & 0.852 & -0.966 & 0.968 & 1.137 & $< 0.001$ \\ 
 $\mu_{3}$ & $\theta_{1}$ & 0.002 & 0.990 & 0.009 & 0.794 & 0.000 & 0.838 & 0.013 & 0.545 & 0.000 & 0.990 & 0.003 & 0.342 \\ 
 $\mu_{3}$ & $\theta_{2}$ & 0.002 & 0.986 & 0.009 & 0.810 & 0.000 & 0.912 & 0.021 & 0.553 & 0.050 & 0.962 & 0.022 & 1.000 \\ 
 $\mu_{3}$ & $\theta_{3}$ & -0.041 & 0.960 & 0.027 & $< 0.001$ & -0.043 & 0.952 & 0.048 & < 0.001 & 0.058 & 0.966 & 0.034 & 1.000 \\ 
 $\mu_{3}$ & $\theta_{4}$ & -1.138 & 0.954 & 1.387 & 0.001 & 1.975 & 0.954 & 2.892 & 0.997 & -0.764 & 0.956 & 1.180 & 0.005 \\ 
 \bottomrule
    \end{tabular}
    \caption{Simulation results for the settings of the non-linear conditional mean function. All twelve combinations of the three baseline risk functions (\ref{eq:mu-sim2}) and four ITE functions (\ref{eq:theta-sim2}) are considered. Linear regression, Lasso, and boosting methods are compared for sample sizes of $n = 500$ and $n=2000$. The average value of the predictive estimand $\theta_{XY}$ across simulation replications is reported, and the coverage
and average width of the nested cross-validation based confidence intervals for $\theta_{XY}$ are shown for each method and setting. 
Median one-sided h-values across simulation replications are displayed for each method and simulation setting.  }
    \label{tab:sim2}
\end{table}

\subsection{Non-linear conditional mean functions with modified outcome methods}\label{sim3}

In this simulation study, we consider the same outcome model (\ref{eq:sim2}), baseline risk functions (\ref{eq:mu-sim2}), and ITE functions (\ref{eq:theta-sim2}) used in the simulation study of Section \ref{sim2}. In this simulation study, we examine the use of different modified outcome methods and the operating characteristics of the confidence intervals for the modified predictive estimands $\tilde{\theta}_{XY}$
defined in (\ref{eq:modified_estimand}). We considered the following four methods for this simulation study: linear regression
with modified outcomes $W_{i}$, $L_{1}$-penalized regression (Lasso) 
with modified outcomes $W_{i}$, boosting with modified outcomes $W_{i}$, and $L_{2}$-penalized regression (ridge regression) with modified outcomes $W_{i}$.

\begin{table}[ht]
\scriptsize
    \centering
    \begin{tabular}{lll|rcrr|rcrr}
   \multicolumn{2}{c}{} & \multicolumn{4}{c}{Method = linear}& \multicolumn{4}{c}{Method = Lasso}\\
   \midrule
$n$ & $\mu(\cdot)$ & $\theta(\cdot)$ &$\tilde{\theta}_{XY}$ & Coverage & CI & Median & $\tilde{\theta}_{XY}$ & Coverage & CI & Median\\
 & & & & Proportion & Width & h-value & & Proportion &  Width& h-value\\
 \midrule
500 & $\mu_{1}$ & $\theta_{1}$ & 13.91 & 0.970 & 80.04 & 0.785 & 0.00 & 0.932 & 0.12 & 0.498 \\ 
   & $\mu_{1}$ & $\theta_{2}$ & 14.79 & 0.960 & 80.67 & 0.798 & 0.00 & 0.950 & 0.18 & 0.489 \\ 
   & $\mu_{1}$ & $\theta_{3}$ & 16.18 & 0.954 & 88.91 & 0.783 & 0.00 & 0.942 & 0.18 & 0.491 \\ 
   & $\mu_{1}$ & $\theta_{4}$ & 30.26 & 0.966 & 195.00 & 0.734 & 0.00 & 0.942 & 0.35 & 0.489 \\ 
   & $\mu_{2}$ & $\theta_{1}$ & 0.17 & 0.974 & 0.92 & 0.807 & 0.00 & 0.952 & 0.00 & 0.496 \\ 
   & $\mu_{2}$ & $\theta_{2}$ & 0.22 & 0.964 & 1.14 & 0.806 & 0.00 & 0.956 & 0.00 & 0.497 \\ 
   & $\mu_{2}$ & $\theta_{3}$ & 0.25 & 0.958 & 2.28 & 0.688 & 0.00 & 0.964 & 0.03 & 0.490 \\ 
   & $\mu_{2}$ & $\theta_{4}$ & 1.13 & 0.952 & 36.78 & 0.588 & -0.01 & 0.892 & 0.25 & 0.497 \\ 
   & $\mu_{3}$ & $\theta_{1}$ & 1.56 & 0.960 & 7.76 & 0.793 & 0.00 & 0.950 & 0.11 & 0.479 \\ 
   & $\mu_{3}$ & $\theta_{2}$ & 1.23 & 0.960 & 6.45 & 0.792 & 0.00 & 0.960 & 0.06 & 0.485 \\ 
   & $\mu_{3}$ & $\theta_{3}$ & 0.63 & 0.956 & 4.05 & 0.729 & 0.00 & 0.956 & 0.06 & 0.492 \\ 
   & $\mu_{3}$ & $\theta_{4}$ & -0.45 & 0.954 & 30.38 & 0.498 & -0.03 & 0.892 & 0.28 & 0.507 \\ 
  \midrule
 2000 & $\mu_{1}$ & $\theta_{1}$ & 3.36 & 0.956 & 20.28 & 0.788 & 0.00 & 0.936 & 0.00 & 0.513 \\ 
   & $\mu_{1}$ & $\theta_{2}$ & 3.60 & 0.984 & 20.77 & 0.784 & 0.00 & 0.944 & 0.00 & 0.512 \\ 
   & $\mu_{1}$ & $\theta_{3}$ & 3.80 & 0.958 & 22.41 & 0.780 & 0.00 & 0.952 & 0.00 & 0.498 \\ 
   & $\mu_{1}$ & $\theta_{4}$ & 3.60 & 0.954 & 52.82 & 0.604 & 0.00 & 0.954 & 0.01 & 0.505 \\ 
   & $\mu_{2}$ & $\theta_{1}$ & 0.04 & 0.954 & 0.22 & 0.776 & 0.00 & 0.942 & 0.00 & 0.502 \\ 
   & $\mu_{2}$ & $\theta_{2}$ & 0.05 & 0.960 & 0.28 & 0.809 & 0.00 & 0.968 & 0.00 & 0.505 \\ 
   & $\mu_{2}$ & $\theta_{3}$ & -0.06 & 0.960 & 0.66 & 0.341 & 0.00 & 0.888 & 0.01 & 0.523 \\ 
   & $\mu_{2}$ & $\theta_{4}$ & -3.52 & 0.958 & 12.56 & 0.142 & -0.01 & 0.858 & 0.08 & 0.528 \\ 
   & $\mu_{3}$ & $\theta_{1}$ & 0.37 & 0.966 & 1.89 & 0.792 & 0.00 & 0.960 & 0.00 & 0.483 \\ 
   & $\mu_{3}$ & $\theta_{2}$ & 0.32 & 0.962 & 1.53 & 0.786 & 0.00 & 0.966 & 0.00 & 0.503 \\ 
   & $\mu_{3}$ & $\theta_{3}$ & 0.02 & 0.958 & 1.11 & 0.552 & 0.00 & 0.870 & 0.01 & 0.515 \\ 
   & $\mu_{3}$ & $\theta_{4}$ & -3.73 & 0.964 & 10.93 & 0.074 & -0.01 & 0.868 & 0.12 & 0.531 \\ 
 \midrule
   \multicolumn{2}{c}{} & \multicolumn{4}{c}{Method = boosting}& \multicolumn{4}{c}{Method = ridge regression}\\
 \midrule
 500 & $\mu_{1}$ & $\theta_{1}$ & 9.75 & 0.970 & 67.56 & 0.764 & 0.00 & 0.946 & 0.05 & 0.487 \\ 
   & $\mu_{1}$ & $\theta_{2}$ & 10.43 & 0.962 & 67.06 & 0.755 & 0.00 & 0.930 & 0.05 & 0.486 \\ 
   & $\mu_{1}$ & $\theta_{3}$ & 11.29 & 0.956 & 74.01 & 0.747 & 0.00 & 0.946 & 0.11 & 0.495 \\ 
   & $\mu_{1}$ & $\theta_{4}$ & 20.21 & 0.958 & 162.20 & 0.693 & 0.00 & 0.932 & 0.18 & 0.510 \\ 
   & $\mu_{2}$ & $\theta_{1}$ & 0.11 & 0.968 & 0.74 & 0.763 & 0.00 & 0.946 & 0.00 & 0.480 \\ 
   & $\mu_{2}$ & $\theta_{2}$ & 0.16 & 0.954 & 0.92 & 0.749 & 0.00 & 0.944 & 0.00 & 0.501 \\ 
   & $\mu_{2}$ & $\theta_{3}$ & 0.12 & 0.970 & 1.88 & 0.627 & 0.00 & 0.938 & 0.01 & 0.499 \\ 
   & $\mu_{2}$ & $\theta_{4}$ & -0.62 & 0.968 & 32.61 & 0.474 & -0.01 & 0.882 & 0.11 & 0.516 \\ 
   & $\mu_{3}$ & $\theta_{1}$ & 1.09 & 0.962 & 6.45 & 0.747 & 0.00 & 0.944 & 0.05 & 0.481 \\ 
   & $\mu_{3}$ & $\theta_{2}$ & 0.86 & 0.966 & 5.26 & 0.747 & 0.00 & 0.962 & 0.03 & 0.488 \\ 
   & $\mu_{3}$ & $\theta_{3}$ & 0.38 & 0.964 & 3.42 & 0.671 & 0.00 & 0.970 & 0.03 & 0.489 \\ 
   & $\mu_{3}$ & $\theta_{4}$ & -1.63 & 0.948 & 26.71 & 0.441 & 0.00 & 0.888 & 0.12 & 0.517 \\ 
  \midrule
  2000 & $\mu_{1}$ & $\theta_{1}$ & 2.40 & 0.962 & 16.53 & 0.774 & 0.00 & 0.950 & 0.00 & 0.502 \\ 
   & $\mu_{1}$ & $\theta_{2}$ & 2.57 & 0.968 & 17.27 & 0.748 & 0.00 & 0.940 & 0.00 & 0.499 \\ 
   & $\mu_{1}$ & $\theta_{3}$ & 2.65 & 0.970 & 19.13 & 0.737 & 0.00 & 0.950 & 0.00 & 0.486 \\ 
   & $\mu_{1}$ & $\theta_{4}$ & 1.19 & 0.952 & 44.86 & 0.578 & 0.00 & 0.936 & 0.00 & 0.514 \\ 
   & $\mu_{2}$ & $\theta_{1}$ & 0.03 & 0.960 & 0.19 & 0.755 & 0.00 & 0.934 & 0.00 & 0.499 \\ 
   & $\mu_{2}$ & $\theta_{2}$ & 0.04 & 0.960 & 0.23 & 0.765 & 0.00 & 0.972 & 0.00 & 0.497 \\ 
   & $\mu_{2}$ & $\theta_{3}$ & -0.10 & 0.948 & 0.56 & 0.233 & 0.00 & 0.862 & 0.01 & 0.516 \\ 
   & $\mu_{2}$ & $\theta_{4}$ & -3.93 & 0.952 & 11.18 & 0.085 & 0.00 & 0.866 & 0.03 & 0.511 \\ 
   & $\mu_{3}$ & $\theta_{1}$ & 0.26 & 0.970 & 1.58 & 0.781 & 0.00 & 0.948 & 0.00 & 0.496 \\ 
   & $\mu_{3}$ & $\theta_{2}$ & 0.22 & 0.936 & 1.27 & 0.749 & 0.00 & 0.956 & 0.00 & 0.492 \\ 
   & $\mu_{3}$ & $\theta_{3}$ & -0.04 & 0.968 & 0.94 & 0.465 & 0.00 & 0.892 & 0.00 & 0.515 \\ 
   & $\mu_{3}$ & $\theta_{4}$ & -4.05 & 0.954 & 10.22 & 0.050 & -0.01 & 0.884 & 0.06 & 0.528 \\
 \bottomrule
    \end{tabular}
    \caption{Simulation results for \textbf{modified outcome methods} using the 12 settings of the non-linear conditional mean function. All twelve combinations of the three baseline risk functions (\ref{eq:mu-sim2}) and four ITE functions (\ref{eq:theta-sim2}) are considered. Linear regression, Lasso, and boosting applied to modified outcomes (\ref{eq:modfied_outcomes_definition}) are compared.
    These methods are evaluated for sample sizes of $n = 500$ and $n=2000$. The average value of the modified predictive estimand $\tilde{\theta}_{XY}$ across simulation replications is reported, and the coverage
and average width of the nested cross-validation based confidence intervals for $\theta_{XY}$ are shown for each method and setting. 
Median one-sided h-values across simulation replications are displayed for each method and simulation setting.}
    \label{tab:sim3}
\end{table}

Table \ref{tab:sim3} shows the mean values of the modified outcomes-based predictive difference quantity $\tilde{\theta}_{XY}$ for the 12 possible settings of the baseline risk and ITE functions.
Similar to the simulations from Section \ref{sim2}, the predictive estimand of interest shows that, when
the sample size is $n = 500$, the restricted estimators have at least as good or better performance than 
the unrestricted estimators across all methods,
and hence, median h-values in the $n = 500$ simulation runs are nowhere close to zero. The lack of any improvement in predictive performance when $n = 500$ is likely due to the much higher variation of the modified outcomes
which makes it difficult for most methods to estimate the ITE function with any accuracy.
The main exception to this is
boosting with baseline risk function $\mu_{3}$ and ITE function $\theta_{4}$, where the
average difference in the predictive performance of the unrestricted estimator versus
the unrestricted estimator $\tau_{MO}^{*}$ is $1.63$.

For the simulations with the larger sample size of $n = 2000$, many of the settings
with HTE have an estimand which is still nonnegative or close to zero if negative. 
This is particularly true for the settings with baseline risk function $\mu_{1}(\mathbf{x})$,
which is likely attributable to the considerable variation of the baseline risk function 
in this setting. It is only the settings with baseline risk $\mu_{2}(\mathbf{x})$ or $\mu_{3}(\mathbf{x})$
and ITE function $\theta_{4}(\mathbf{x})$ where $\tilde{\theta}_{XY}$ in the boosting
and linear regression methods clearly favors the unrestricted estimator. Even in these settings,
the variation in the modified outcomes $W_{i}$ is such that only boosting
in the setting with $\mu_{3}(\mathbf{x})$ and $\theta_{4}(\mathbf{x})$
has enough power to ensure that h-values are frequently less than $0.05$.
Overall, the results from these simulations highlight the need for quite large
sample sizes when using direct modified outcomes defined in (\ref{eq:modfied_outcomes_definition}). 
While using modified outcomes does eliminate the need for modeling and fitting  
the full conditional mean function $f(A, \mathbf{x})$, the 
increase in variance can often lead to low power to confidently
capture any predictive gains of the unrestricted estimator versus
the restricted estimator.

Across all settings, the coverage proportions are quite close to the nominal $95\%$ level
with the exception of a few cases where the Lasso has under $90\%$ coverage. The most notable thing 
about the confidence intervals for the Lasso and ridge regression methods is that they are extremely narrow. 
This arises from the fact that Lasso typically selects none of the covariates
which implies that the unrestricted and restricted estimators are exactly the same, and hence
the values of the difference loss (\ref{eq:main_loss_function}) are almost always zero 
when evaluated in the nested cross-validation procedure. Similarly, ridge regression usually 
shrinks all regression coefficients very close to zero causing the unrestricted and 
restricted estimators to be very similar. Note that both Lasso and ridge regression
use the default cross-validation-based procedures in the \textit{glmnet}
package \cite{friedman2010} to select the hyperparameters, and an alternative hyperparameter
selection procedure specifically targeted for modified outcomes could yield
different results.

\section{Application: The IHDP Trial}\label{sec:App}

This section illustrates the use of our methodology by analyzing outcomes from the Infant Health and Development Program (IHDP)\cite{brooks1992, hill2011}. This program was a randomized trial that began in 1985 and focused on evaluating different interventions for low-birthweight premature infants. The treatment group received high-quality child care and home visits from a trained provider at eight clinical sites. Initially, the study included 985 individuals, and after cleaning the dataset by removing observations that had missing outcome values, we had 908 observations in total.
The outcome variable is the result of an IQ assessment of the child at 36 months. To remove predictors with very little variation or association with the outcome, we only included the 10 variables most correlated with the outcome variable, along with the mother’s age, as predictors. These selected predictors include the mother's Peabody Picture Vocabulary Test (PPVT) measured 1 year post-birth, the mother’s ethnicity, measures of maternal education level, marital status at birth, birth order, number of children, and clinical site number.   

\begin{figure}[ht]
    \centering
    \includegraphics[width=1\textwidth]{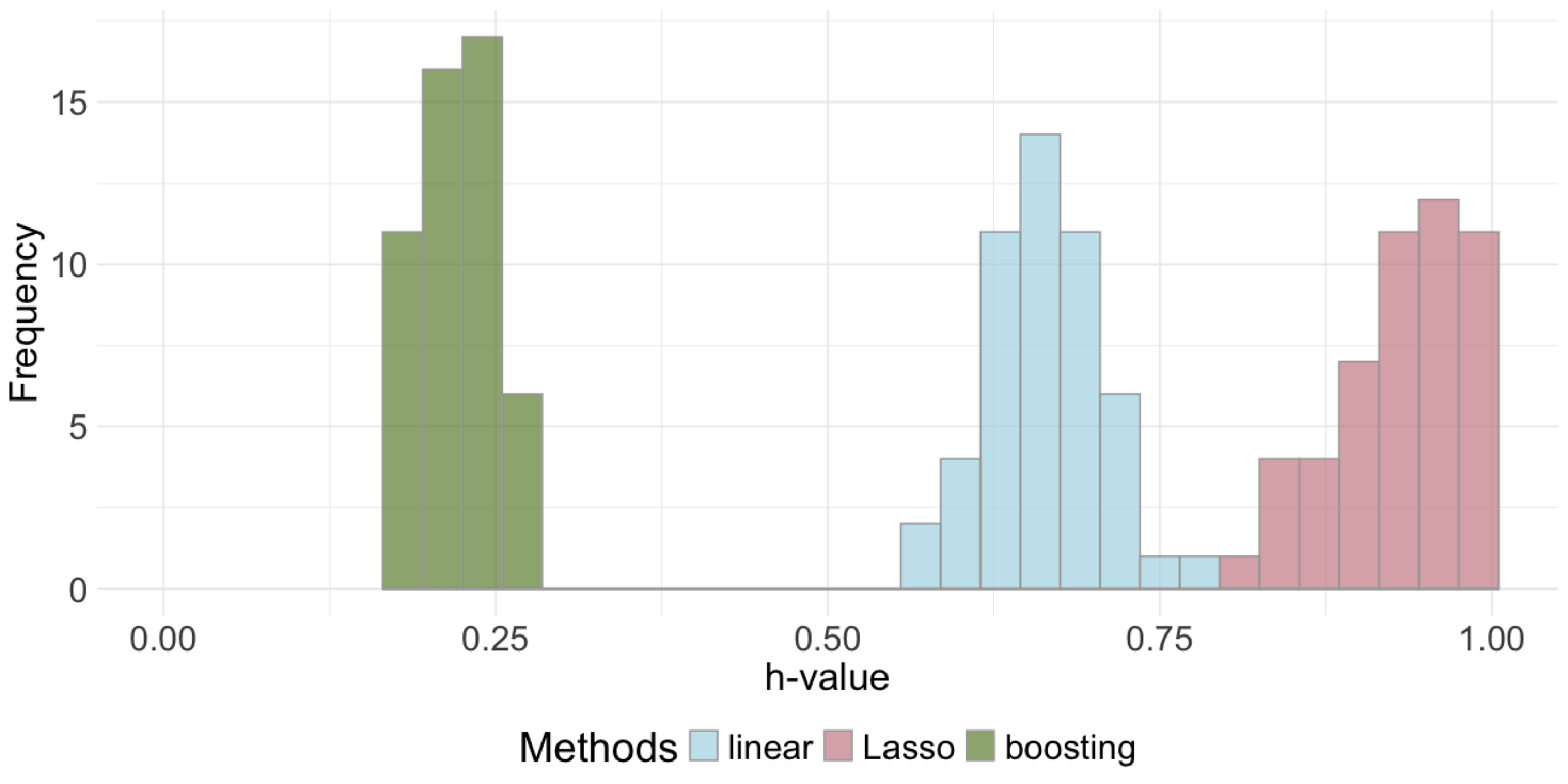}
    \caption{Histogram of h-values computed using linear regression, Lasso, and boosting approaches for the IHDP dataset. For each method, 50 h-values were computed. To compute each h-value, five-fold nested cross-validation and 50 repetitions were used. }
    \label{fig:histogram-hvalues}
\end{figure}

Figure \ref{fig:histogram-hvalues} displays histograms of h-values obtained from linear, Lasso, and boosting procedures. Each h-value was computed using five-fold nested cross-validation with 50 repetitions of the nested cross-validation procedure, and for each of the three methods, we computed 50 separate h-values. Each histogram in Figure \ref{fig:histogram-hvalues} characterizes the distribution of the $50$ h-values for that method. As shown in these histograms, 
there is still modest variation in the h-values when re-computing h-values on the same dataset, even when using $50$ nested cross-validation repetitions. If one wants minimal sampling-based variation in the h-values, one may need to use
a very large number of repetitions such as $500$ or more.
Despite this variation in h-values, the histograms reveal quite distinct clusterings of h-values across methods used. Linear regression predominantly produces h-values in the ranges of $0.5$ and $0.75$ while h-values from Lasso cluster closer to one with a modal h-value between $0.9$ and 1.0. The boosting method has h-values that range from roughly $0.2$ to $0.25$. Thus, for each of the three methods
considered, there is no strong evidence suggesting that modeling HTE leads to improved predictive performance.

\begin{figure}[ht]
    \centering
    \includegraphics[width=16cm, height=12cm]{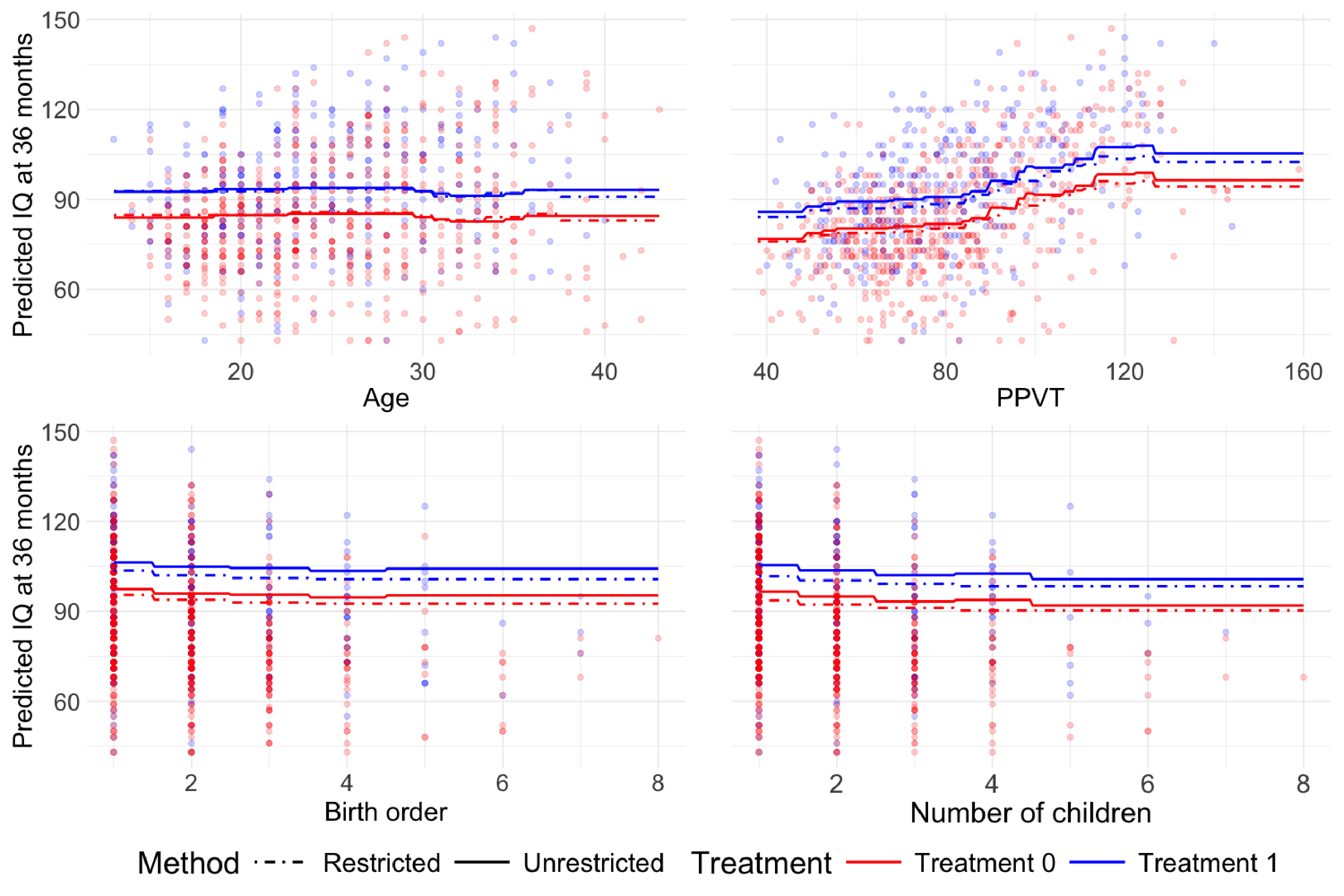}
    \caption{Partial dependence plots from the boosting approach for age, PPTV, birth order, and number of children variables.}
    \label{fig:par-dep}
\end{figure}

Plotting partial dependence functions \cite{friedman2001} is a useful way to explore how the estimated conditional mean function is related to the variation in certain covariates. 
For the conditional mean function $f(A, \mathbf{x})$, we define the partial dependence function for covariate $k$ within treatment arm $a$ at covariate value $u$, for the unrestricted estimator, as 
\begin{equation}
\rho_{k}(a, u) = \frac{1}{n}\sum_{i=1}^{n} \hat{f}\big(a, (u, \mathbf{x}_{i,-k}); \mathcal{D}\big),
\label{eq:partial_dep}
\end{equation}
and the partial dependence function for the restricted estimator is defined similarly.
In (\ref{eq:partial_dep}), the notation $(\mathbf{x}_{i,-k}, u)$ denotes the vector where $x_{ik} = u$ 
and the remaining covariates are set equal to their observed values.
Figure \ref{fig:par-dep} presents the boosting partial dependence plots for the following variables: maternal age, PPVT, birth order, and the number of children. Within each treatment group, estimated measured IQ at 36 months changes very little according to maternal age, with a slight decrease between ages 30 and 35. The partial dependence functions for birth order and the number of children also show very little variation but do exhibit a modest decreasing trend. 
The PPVT partial dependence plot shows a more clear, positive association between this variable and the predicted IQs
with the partial dependence functions exhibiting some nonlinearity as the PPVT moves into the higher values of PPVT,
but as with the other variables shown in the plot, the difference between the treatment-specific 
partial dependence functions $\rho_{k}(1, u)$ and $\rho_{k}(0, u)$ is quite constant across all ranges of PPVT.

The differences between the partial dependence functions $\rho_{k}(1, u)$ and $\rho_{k}(0, u)$ in Figure \ref{fig:par-dep} do provide an indication of why most of the boosting-based h-values exceeded $0.2$. While these partial dependence plots do not contain
all the relevant information about the differences between the unrestricted and restricted estimators of the conditional mean function, the 
small variation in the differences between $\rho_{k}(1, u)$ and $\rho_{k}(0, u)$ suggests there will be relatively modest differences between the predictions of the unrestricted and restricted estimators.
For each variable displayed, the difference between the treatment-arm partial dependence function and the control-arm partial dependence function is relatively constant over the range of the variable considered, which implies that the 
average treatment effect estimate (averaging over covariates other than $k$) is constant. 
In addition to this, the differences between the partial dependence functions of the unrestricted and restricted estimators are quite small. 
Hence, based on these partial dependence plots alone, one would not expect the predictive performance of the unrestricted estimator to be substantially different from that of the restricted estimator.


\section{Discussion}

We have proposed a procedure for evaluating the predictive benefit of including HTE 
when modeling outcomes in a randomized study. To do this, our approach compares 
the out-of-sample prediction from a model that allows for treatment effect variation
with a similar version of that model that is restricted to have constant treatment effect.
We constructed confidence intervals for the predictive difference estimands $\theta_{XY}$ and $\tilde{\theta}_{XY}$
and their associated h-values using a particular nested cross-validation procedure.
However, our framework for evaluating the predictive benefit of incorporating HTE into a statistical
model is more general and does not rely on a particular method for constructing confidence intervals
for $\theta_{XY}$ or $\tilde{\theta}_{XY}$. Rather, the essential feature of our approach
is uncertainty quantification for the difference in out-of-sample performance between the 
unrestricted estimator $\hat{f}(A, \mathbf{x}; \mathcal{D})$ and the restricted estimator 
$\hat{g}(A, \mathbf{x}; \mathcal{D})$, and one could certainly explore alternative approaches for constructing
uncertainty intervals for $\theta_{XY}$ or $\tilde{\theta}_{XY}$. For example, if one has found that an alternative subsampling
procedure tends to generate confidence intervals for $\theta_{XY}$ with improved coverage
or narrower confidence intervals with similar coverage, then that procedure can be used instead 
of the nested cross-validation procedure to generate 
an uncertainty interval for $\theta_{XY}$.

In our simulation studies, we found that the nested cross-validation procedure \cite{bates2024}
usually provides close to nominal coverage -- particularly for linear regression, 
penalized linear regression, and boosting methods. For random forests
and Bayesian additive regression trees (BART) \cite{chipman2010bart},
we have observed a few simulation settings where this nested cross-validation
procedure yields coverage proportions which are somewhat less
than the nominal coverage (see Appendix C) though
the coverage performance of random forest and BART was strong for most simulation settings.
To improve the robustness of coverage for such nonlinear estimators,
it may be worth exploring the use of alternative standard errors
for cross-validation-type estimators or investigating the use of automatic
variance-stabilizing transformation procedures \cite{tibshirani1988},
as it may often be the case that the values of the hold-out losses and 
their standard errors have a stronger association for nonlinear
methods such as random forests.

In our proposed method, we focused on the setting where outcomes are continuous and the
natural notion of a treatment effect is the expected difference between the outcome
under the active treatment versus under the control. However, for different outcome types
such as binary or survival outcomes, one may need to modify the definition of the
``restricted estimator'' appropriately to reflect the scale on which one wants to model
the relationship between the treatment effect and covariates.
For example, consider computing a restricted estimator $\hat{g}(A, \mathbf{x}; \mathcal{D})$ when 
one has binary outcomes, and one wants to model treatment effect variation on the log-odds
scale rather than the risk difference scale. In this case, one would want the restricted estimator differences
$\hat{g}(1, \mathbf{x}; \mathcal{D}) - \hat{g}(0, \mathbf{x}; \mathcal{D})$ to be constant as a function
of $\mathbf{x}$ only if $\hat{g}(A, \mathbf{x}; \mathcal{D})$ represents a log-odds. 
The challenge in this setting would be how to construct a restricted estimator $\hat{g}(A, \mathbf{x}; \mathcal{D})$
that satisfies this property using the shifted outcomes approach described in Section \ref{sec:Method}. 
One possibility to address this challenge would be to, rather than shift the outcomes,
randomly switch some of the labels in the active treatment arm in such a way that the success probabilities
approximately match those in the control arm. 
Exploring and implementing such an approach is beyond the scope of the present work,
but investigating similar ideas would be an interesting topic for future work.

Our predictive estimand compares the performance of an unrestricted estimator that allows
for treatment effect variation with a restricted estimator that we characterized as
the closest constant-treatment-effect version of the unrestricted estimator. 
We did this to more clearly separate the impact of incorporating
HTE from predictive gains that may arise from using different estimation
methodologies for the unrestricted and restricted estimators.
However, one could also compare unrestricted and restricted estimators
constructed using different statistical methods if such a comparison was
considered to be more relevant in certain settings. For example, one could 
compare the performance of an unrestricted random forest procedure with a restricted
tree-boosting-based estimator that is computed using the shifted-outcomes approach 
described in Section \ref{sec:Method}. In this case, the interpretation of 
the predictive estimand would be the out-of-sample performance improvement
resulting from using a random forest approach that allows for HTE 
compared to a tree-boosting approach that does not allow for treatment effect variation.

The description of our method and simulation studies focused 
on the setting where study participants are assigned randomly to 
two treatment arms. However, large observational studies are increasingly being
utilized as a tool to identify HTE in real-world settings and exploring the use of our framework to observational 
settings would certainly be worthwhile. Our approach could easily be adapted to observational settings
if the primary focus were on the ITE-targeted predictive estimand $\tilde{\theta}_{XY}$ described 
in Section \ref{sec:ModifiedOM}. Specifically, if one assumes that enough relevant covariates
have been collected to ensure no unmeasured confounding, one would create
modified outcomes using estimated propensity scores rather than design-determined
probabilities of treatment assignment.
Using these modified outcomes together with loss function (\ref{eq:modified_loss_function})
would still guarantee that the predictive estimand $\tilde{\theta}_{XY}$
has the same interpretation as a difference in predictive performance of the unrestricted
ITE estimate versus the restricted ITE estimate.

\subsection*{Supplementary Information}
The IHDP dataset used in this paper is publicly available at: \\
\url{https://github.com/mahsaashouri/HTE-Model-Comparison}.

\noindent
R code used to reproduce simulation results and analysis of the IHDP dataset is also available at: \url{https://github.com/mahsaashouri/HTE-Model-Comparison}.

\bibliographystyle{agsm}
\bibliography{ref}

\appendix

\section{Expression for Conditional Expectation of Modified Loss}
In Section 3 of the main manuscript, we defined the modified loss $\tilde{\ell}$ as
\begin{equation}
\tilde{\ell}\Big( \hat{\theta}(\mathbf{x}_{n+1}; \mathcal{D}), \tau_{MO}^{*}, W_{n+1} \Big)
= \Big( W_{n+1} - \hat{\theta}(\mathbf{x}_{n+1}; \mathcal{D}) \Big)^{2} 
- \Big( W_{n+1} - \tau_{MO}^{*} \Big)^{2},
\end{equation}
and the modified predictive estimand $\tilde{\theta}_{XY}$ as 
\begin{equation}
\tilde{\theta}_{XY} = E\Big\{ \tilde{\ell}\Big( \hat{\theta}(\mathbf{x}_{n+1}; \mathcal{D}), \tau_{MO}^{*}, W_{n+1} \Big) \Big| \mathcal{D} \Big\}. \nonumber 
\end{equation}

\medskip

\noindent
To better interpret $\tilde{\theta}_{XY}$, one can first note
\begin{multline}
E\Big\{ \Big( W_{n+1} - \hat{\theta}(\mathbf{x}_{n+1}; \mathcal{D}) \Big)^{2} \Big| \mathcal{D}, \mathbf{x}_{n+1} \Big\}
= E\Big\{ \Big( W_{n+1} - \theta(\mathbf{x}_{n+1}) \Big)^{2} \Big| \mathcal{D}, \mathbf{x}_{n+1} \Big\} \\
+ 2E\Big\{ \Big( W_{n+1} - \theta(\mathbf{x}_{n+1}) \Big) \Big| \mathcal{D}, \mathbf{x}_{n+1} \Big\}E\Big\{ \Big( \theta(\mathbf{x}_{n+1}) - \hat{\theta}(\mathbf{x}_{n+1}; \mathcal{D}) \Big) \Big| \mathcal{D}, \mathbf{x}_{n+1} \Big\} \\
+ E\Big\{ \Big( \theta(\mathbf{x}_{n+1}) - \theta(\mathbf{x}_{n+1}) \Big)^{2} \Big| \mathcal{D}, \mathbf{x}_{n+1} \Big\} \\
=  E\Big\{ \Big( W_{n+1} - \theta(\mathbf{x}_{n+1}) \Big)^{2} \Big| \mathcal{D}, \mathbf{x}_{n+1} \Big\} + E\Big\{ \Big( \theta(\mathbf{x}_{n+1}) - \hat{\theta}(\mathbf{x}_{n+1}; \mathcal{D}) \Big)^{2} \Big| \mathcal{D}, \mathbf{x}_{n+1} \Big\},
\label{eq:wsqderror}
\end{multline}
where the first equality follows from the fact that $(W_{n+1} - \theta(\mathbf{x}_{n+1}))$ and
$(\theta(\mathbf{x}_{n+1}) - \hat{\theta}(\mathbf{x}_{n+1}; \mathcal{D}))$ are independent given $(\mathcal{D}, \mathbf{x}_{n+1})$
and where the second equality follows from the fact that $E\{ W_{n+1} |\mathcal{D}, \mathbf{x}_{n+1}\} = \theta(\mathbf{x}_{n+1})$.
An identical argument will show that
\begin{eqnarray}
E\Big\{ \Big( W_{n+1} - \tau_{MO}^{*} \Big)^{2} \Big| \mathcal{D}, \mathbf{x}_{n+1} \Big\}
&=&  E\Big\{ \Big( W_{n+1} - \theta(\mathbf{x}_{n+1}) \Big)^{2} \Big| \mathcal{D}, \mathbf{x}_{n+1} \Big\} \nonumber \\
&+& E\Big\{ \Big( \theta(\mathbf{x}_{n+1}) - \tau_{MO}^{*} \Big)^{2} \Big| \mathcal{D}, \mathbf{x}_{n+1} \Big\}.
\label{eq:tausqderror}
\end{eqnarray}
By combining (\ref{eq:wsqderror}) and (\ref{eq:tausqderror}), we have that
\begin{eqnarray}
&& E\Bigg\{ \tilde{\ell}\Big( \hat{\theta}(\mathbf{x}_{n+1}; \mathcal{D}), \tau_{MO}^{*}, W_{n+1} \Big) \Big| \mathcal{D}, \mathbf{x}_{n+1} \Bigg\} \nonumber \\
&=& E\Bigg\{ \Big( W_{n+1} - \hat{\theta}(\mathbf{x}_{n+1}; \mathcal{D}) \Big)^{2} \Big| \mathcal{D}, \mathbf{x}_{n+1} \Bigg\}
- E\Bigg\{ \Big( W_{n+1} - \tau_{MO}^{*} \Big)^{2} \Big| \mathcal{D}, \mathbf{x}_{n+1} \Bigg\} \nonumber \\
&=& E\Big\{ \Big( \theta(\mathbf{x}_{n+1}) - \hat{\theta}(\mathbf{x}_{n+1}; \mathcal{D}) \Big)^{2} \Big| \mathcal{D}, \mathbf{x}_{n+1} \Big\}
- E\Big\{ \Big( \theta(\mathbf{x}_{n+1}) - \tau_{MO}^{*} \Big)^{2} \Big| \mathcal{D}, \mathbf{x}_{n+1} \Big\}, \nonumber
\end{eqnarray}
and hence,
\begin{multline}
E\Bigg\{ \tilde{\ell}\Big( \hat{\theta}(\mathbf{x}_{n+1}; \mathcal{D}), \tau_{MO}^{*}, W_{n+1} \Big) \Big| \mathcal{D} \Bigg\}
=  E\Bigg[ E\Big\{ \tilde{\ell}\Big( \hat{\theta}(\mathbf{x}_{n+1}; \mathcal{D}), \tau_{MO}^{*}, W_{n+1} \Big) \Big| \mathcal{D}, \mathbf{x}_{n+1} \Big\} \Bigg| \mathcal{D}\Bigg] \\
= E\Big\{ \Big( \theta(\mathbf{x}_{n+1}) - \hat{\theta}(\mathbf{x}_{n+1}; \mathcal{D}) \Big)^{2} \Big| \mathcal{D} \Big\}
- E\Big\{ \Big( \theta(\mathbf{x}_{n+1}) - \tau_{MO}^{*} \Big)^{2} \Big| \mathcal{D} \Big\}. \nonumber
\end{multline}

\section{Restricted Estimator in the Linear Model with a Treatment-Scalar Covariate Interaction Example}
This section concerns the restricted treatment effect estimate $\tau^{*}$ in the linear model with a scalar
covariate example described in Section 2.2 of the main manuscript.

We will first define useful notation for this example. We define $\bar{Y}$, $\bar{x}$, $\bar{A}$,
$\hat{\mu}_{xy}$, $\hat{\mu}_{ay}$, $\hat{\mu}_{xx}$ as
$\bar{Y} = n^{-1}\sum_{i=1}^{n} Y_{i}$, $\bar{x} = n^{-1}\sum_{i=1}^{n} x_{i}$, $\bar{A} = n^{-1}\sum_{i=1}^{n} A_{i}$,
$\hat{\mu}_{xy} = n^{-1}\sum_{i=1}^{n} x_{i}Y_{i}$, $\hat{\mu}_{ay} = n^{-1}\sum_{i=1}^{n} A_{i}Y_{i}$, 
$\hat{\mu}_{xx} = n^{-1}\sum_{i=1}^{n}x_{i}^{2}$, and 
we define $s_{x}$ as 
\begin{equation}
s_{x} = \frac{1}{n}\sum_{i=1}^{n} (x_{i} - \bar{x})^{2} = \hat{\mu}_{xx} - \bar{x}^{2}. \nonumber 
\end{equation}
Also, we define $C_{aa}, C_{xa}, C_{xy}, C_{ay}, C_{xy}$ as
\begin{eqnarray}
C_{aa} &=& \sum_{i=1}^{n}(A_{i} - \bar{A})A_{i}\big/\sum_{i=1}^{n}(x_{i} - \bar{x})^{2} = (\bar{A} - \bar{A}^{2})/s_{x} \nonumber \\
C_{xa} &=& \sum_{i=1}^{n}(x_{i} - \bar{x})A_{i}\big/\sum_{i=1}^{n}(x_{i} - \bar{x})^{2} = (\hat{\mu}_{ax} - \bar{A}\bar{x})/s_{x} \nonumber \\
C_{ay} &=& \sum_{i=1}^{n}(Y_{i} - \bar{Y})A_{i}\big/\sum_{i=1}^{n}(x_{i} - \bar{x})^{2} = (\hat{\mu}_{ay} - \bar{A}\bar{Y})/s_{x} \nonumber \\
C_{xy} &=& \sum_{i=1}^{n}(x_{i} - \bar{x})Y_{i}\big/\sum_{i=1}^{n}(x_{i} - \bar{x})^{2} = (\hat{\mu}_{xy} - \bar{Y}\bar{x})/s_{x} \nonumber 
\end{eqnarray}

\bigskip

\noindent
From the description in Section 2 of the main manuscript for finding the restricted treatment effect estimate $\tau^{*}$
for a linear model with a treatment-covariate interaction, it was stated that $\tau^{*}$ is obtained as
the solution of the following minimization problem
\begin{equation}
\tau^{*}
= \argmin_{\tau} \sum_{i=1}^{n} \Big\{ (Y_{i} - \bar{Y}) - C_{xy}(x_{i} - \bar{x}) - \tau[(A_{i} - \bar{A}) - C_{xa}(x_{i} - \bar{x})] \Big\}^{2}. \nonumber 
\end{equation}
A direct calculation shows that $\tau^{*}$ can be expressed as
\begin{equation}
\tau^{*} = \frac{\sum_{i=1}^{n} \{ (Y_{i} - \bar{Y}) - C_{xy}(x_{i} - \bar{x})\}\{ (A_{i} - \bar{A}) - C_{xa}(x_{i} - \bar{x})\}}{\sum_{i=1}^{n} \{ (A_{i} - \bar{A}) - C_{xa}(x_{i} - \bar{x})\}^{2} }. 
\label{eq:tau_form}
\end{equation}
Letting $d^{*} = \sum_{i=1}^{n} \{ (A_{i} - \bar{A}) - C_{xa}(x_{i} - \bar{x})\}^{2}$,
we can further simplify the expression for $\tau^{*}$ in (\ref{eq:tau_form}) to 
\begin{eqnarray}
\tau^{*} &=& \frac{1}{d^{*}}\Big[\sum_{i=1}^{n} (Y_{i} - \bar{Y})(A_{i} - \bar{A}) - C_{xa}\sum_{i=1}^{n}(Y_{i} - \bar{Y})(x_{i} - \bar{x})
- C_{xy}\sum_{i=1}^{n}(x_{i} - \bar{x})(A_{i} - \bar{A}) \nonumber \\
&& + C_{xa}C_{xy}\sum_{i=1}^{n}(x_{i} - \bar{x})^{2} \Big] \nonumber \\
&=& \frac{1}{d^{*}}\Big[nC_{ay}(\hat{\mu}_{xx} - \bar{x}^{2}) - nC_{xa}C_{xy}(\hat{\mu}_{xx} - \bar{x}^{2})
- nC_{xy}C_{xa}(\hat{\mu}_{xx} - \bar{x}^{2}) + nC_{xa}C_{xy}(\hat{\mu}_{xx} - \bar{x}^{2}) \Big]  \nonumber \\
&=& \frac{ns_{x}}{d^{*}}\Big[C_{ay} - C_{xa}C_{xy} \Big]. \nonumber 
\end{eqnarray}
Now to simplify $d^{*}$ in the above, we note that
\begin{eqnarray}
d^{*} &=& \sum_{i=1}^{n} \{ (A_{i} - \bar{A}) - C_{xa}(x_{i} - \bar{x})\}^{2} \nonumber \\
&=& \sum_{i=1}^{n} A_{i}^{2} - n\bar{A}^{2} - 2C_{xa}\sum_{i=1}^{n} \{ (A_{i} - \bar{A})(x_{i} - \bar{x})\}
+ C_{xa}^{2}(n\hat{\mu}_{xx} - n\bar{x}^{2}) \nonumber \\
&=& n\bar{A} - n\bar{A}^{2} - 2nC_{xa}(\hat{\mu}_{ax} - \bar{A}\bar{x}) +  nC_{xa}^{2}(\hat{\mu}_{xx} - \bar{x}^{2}) \nonumber \\
&=& C_{aa}n(\hat{\mu}_{xx} - \bar{x}^{2}) - 2nC_{xa}^{2}(\hat{\mu}_{xx} - \bar{x}^{2}) +  nC_{xa}^{2}(\hat{\mu}_{xx} - \bar{x}^{2}) \nonumber \\
&=& nC_{aa}s_{x} - nC_{xa}^{2}s_{x}. 
\label{eq:d_form}
\end{eqnarray}
Combining (\ref{eq:tau_form}) and (\ref{eq:d_form}), we can express $\tau^{*}$ as: 
\begin{equation}
\tau^{*} = \frac{C_{ay} - C_{xa}C_{xy}}{C_{aa} - C_{xa}^{2}}. \nonumber 
\end{equation}

\bigskip

\bigskip

\noindent
Now, let us consider the linear regression model, 
\begin{equation}
Y_{i} = \beta_{0} + \beta_{1}x_{i} + \tau A_{i} + \varepsilon_{i}, \quad i = 1,\ldots, n. 
\label{eq:model_noint}
\end{equation}
with scalar covariates $x_{1}, \ldots, x_{n}$.
The $n \times 3$ design matrix associated with model (\ref{eq:model_noint}) is
\begin{equation}
\mathbf{X} = \begin{bmatrix} 
1 & x_{1} & A_{1} \\
1 & x_{2} & A_{2} \\
\vdots & \vdots & \vdots \\
1 & x_{n} & A_{n}
\end{bmatrix},
\nonumber 
\end{equation}
and if $\mathbf{Y} = (Y_{1}, \ldots, Y_{n})^{T}$, the
$n \times 1$ vector $\mathbf{X}^{T}\mathbf{Y}$ and $3 \times 3$ matrix $\mathbf{X}^{T}\mathbf{X}$ are given by
\begin{equation}
\mathbf{X}^{T}\mathbf{Y} = n\begin{bmatrix} \bar{Y} \\ \hat{\mu}_{xy} \\ \hat{\mu}_{ay} 
\end{bmatrix}
\qquad\qquad 
\mathbf{X}^{T}\mathbf{X}
= n\begin{bmatrix} 1 & \bar{x} & \bar{A} \\
\bar{x} & \hat{\mu}_{xx} & \hat{\mu}_{ax} \\
\bar{A} & \hat{\mu}_{ax} & \bar{A} 
\end{bmatrix}.
\label{eq:xty}
\end{equation}
The determinant of $n^{-1}\mathbf{X}^{T}\mathbf{X}$ is
\begin{eqnarray}
\det(n^{-1}\mathbf{X}^{T}\mathbf{X}) 
&=& (\hat{\mu}_{xx}\bar{A} - \hat{\mu}_{ax}^{2}) - \bar{x}(\bar{x}\bar{A} - \bar{A}\hat{\mu}_{ax}) + \bar{A}(\bar{x}\hat{\mu}_{ax} - \bar{A}\hat{\mu}_{xx}) \nonumber \\
&=& \hat{\mu}_{xx}\bar{A} - \hat{\mu}_{ax}^{2} - \bar{A}\bar{x}^{2} + 2\bar{A}\bar{x}\hat{\mu}_{ax} -\bar{A}^{2}\hat{\mu}_{xx} \nonumber \\
&=& \hat{\mu}_{xx}(\bar{A} - \bar{A}^{2}) - \hat{\mu}_{ax}^{2} + 2\bar{A}\bar{x}\hat{\mu}_{ax} - \bar{A}^{2}\bar{x}^{2} + \bar{A}^{2}\bar{x}^{2} - \bar{A}\bar{x}^{2} \nonumber \\
&=& \hat{\mu}_{xx}(\bar{A} - \bar{A}^{2}) - (\hat{\mu}_{ax} - \bar{A}\bar{x})^{2} - \bar{x}^{2}(\bar{A} - \bar{A}^{2}) \nonumber \\
&=& (\hat{\mu}_{xx} - \bar{x}^{2})(\bar{A} - \bar{A}^{2}) - (\hat{\mu}_{ax} - \bar{A}\bar{x})^{2} \nonumber \\
&=& s_{x}^{2}(C_{aa} - C_{xa}^{2}), \nonumber 
\end{eqnarray}
and the adjoint of $n^{-1}\mathbf{X}^{T}\mathbf{X}$ is
\begin{eqnarray}
\textrm{adj}(n^{-1}\mathbf{X}^{T}\mathbf{X})
&=& \begin{bmatrix} \hat{\mu}_{xx}\bar{A} - \hat{\mu}_{ax}^{2}  
& \bar{A}\hat{\mu}_{ax} - \bar{x}\bar{A} & \bar{x}\hat{\mu}_{ax} - \bar{A}\hat{\mu}_{xx} \\
\bar{A}\hat{\mu}_{ax} - \bar{x}\bar{A} & \bar{A} - \bar{A}^{2} & \bar{x}\bar{A} - \hat{\mu}_{ax} \\
\bar{x}\hat{\mu}_{ax} - \hat{\mu}_{xx}\bar{A} & \bar{x}\bar{A} - \hat{\mu}_{ax} &  \hat{\mu}_{xx} - \bar{x}^{2}
\end{bmatrix} \nonumber \\
&=& \begin{bmatrix} \hat{\mu}_{xx}\bar{A} - \hat{\mu}_{ax}^{2}  
& \bar{A}\hat{\mu}_{ax} - \bar{x}\bar{A} & \bar{x}\hat{\mu}_{ax} - \bar{A}\hat{\mu}_{xx} \\
\bar{A}\hat{\mu}_{ax} - \bar{x}\bar{A} & C_{aa}s_{x}  & -C_{xa}s_{x} \\
(\bar{x}C_{xa} - \bar{A})s_{x}  & -C_{xa}s_{x} &  s_{x}
\end{bmatrix},
\nonumber 
\end{eqnarray}
where the second equality for the $(3,1)$ entry follows from the fact that
$\bar{x}\hat{\mu}_{ax} - \hat{\mu}_{xx}\bar{A} = \bar{x}\hat{\mu}_{ax} - \bar{x}^{2}\bar{A} + \bar{x}^{2}\bar{A} - \hat{\mu}_{xx}\bar{A}
= \bar{x}(\hat{\mu}_{ax} - \bar{x}\bar{A}) - \bar{A}(\hat{\mu}_{xx} - \bar{x}^{2})$
The inverse $(\mathbf{X}^{T}\mathbf{X})^{-1}
= n^{-1}\textrm{adj}(n^{-1}\mathbf{X}^{T}\mathbf{X})/\det(n^{-1}\mathbf{X}^{T}\mathbf{X})$  of $\mathbf{X}^{T}\mathbf{X}$ is 
then given by
\begin{equation}
(\mathbf{X}^{T}\mathbf{X})^{-1}
= \frac{1}{ns_{x}^{2}(C_{aa} - C_{xa}^{2})}\begin{bmatrix}  \hat{\mu}_{xx}\bar{A} - \hat{\mu}_{ax}^{2}  
& \bar{A}\hat{\mu}_{ax} - \bar{x}\bar{A} & \bar{x}\hat{\mu}_{ax} - \bar{A}\hat{\mu}_{xx} \\
\bar{A}\hat{\mu}_{ax} - \bar{x}\bar{A} & C_{aa}s_{x}  & -C_{xa}s_{x} \\
(\hat{\mu}_{ax}\bar{x} - \hat{\mu}_{x2}\bar{A}) & (\bar{x}\bar{A} - \hat{\mu}_{ax}) &   (\hat{\mu}_{x2} - \bar{x}^{2})
\end{bmatrix}.
\label{eq:xtx_inverse}
\end{equation}
The least-squares estimator $\hat{\tau}_{LS}$ of $\tau$ is the third element of $(\mathbf{X}^{T}\mathbf{X})^{-1}\mathbf{X}^{T}\mathbf{Y}$,
which from (\ref{eq:xty}) and (\ref{eq:xtx_inverse}) is
\begin{eqnarray}
\hat{\tau}_{LS} &=& \frac{1}{ns_{x}^{2}(C_{aa} - C_{xa}^{2})}\Big[n\bar{Y}s_{x}(\bar{x}C_{xa} - \bar{A}) -n\hat{\mu}_{xy}C_{xa}s_{x} + n\hat{\mu}_{ay}s_{x}\Big] \nonumber \\
&=& \frac{1}{ns_{x}^{2}(C_{aa} - C_{xa}^{2})}\Big[ns_{x}C_{xa}\bar{Y}\bar{x} - ns_{x}\bar{Y}\bar{A} -n\hat{\mu}_{xy}C_{xa}s_{x} + n\hat{\mu}_{ay}s_{x}\Big] \nonumber \\
&=& \frac{1}{ns_{x}^{2}(C_{aa} - C_{xa}^{2})}\Big[ns_{x}C_{xa}(\bar{Y}\bar{x} - \hat{\mu}_{xy}) + ns_{x}(\hat{\mu}_{ay} - \bar{Y}\bar{A}) \Big] \nonumber \\
&=& \frac{1}{ns_{x}^{2}(C_{aa} - C_{xa}^{2})}\Big[-ns_{x}^{2}C_{xa}C_{xy} + ns_{x}^{2}C_{ay} \Big] \nonumber \\
&=& \frac{C_{ay} - C_{xa}C_{xy}}{C_{aa} - C_{xa}^{2}} \nonumber \\
&=& \tau^{*}
\end{eqnarray}

\section{Additional Simulation Results}
\begin{figure}[ht]
    \centering
    \includegraphics[width=1\textwidth]{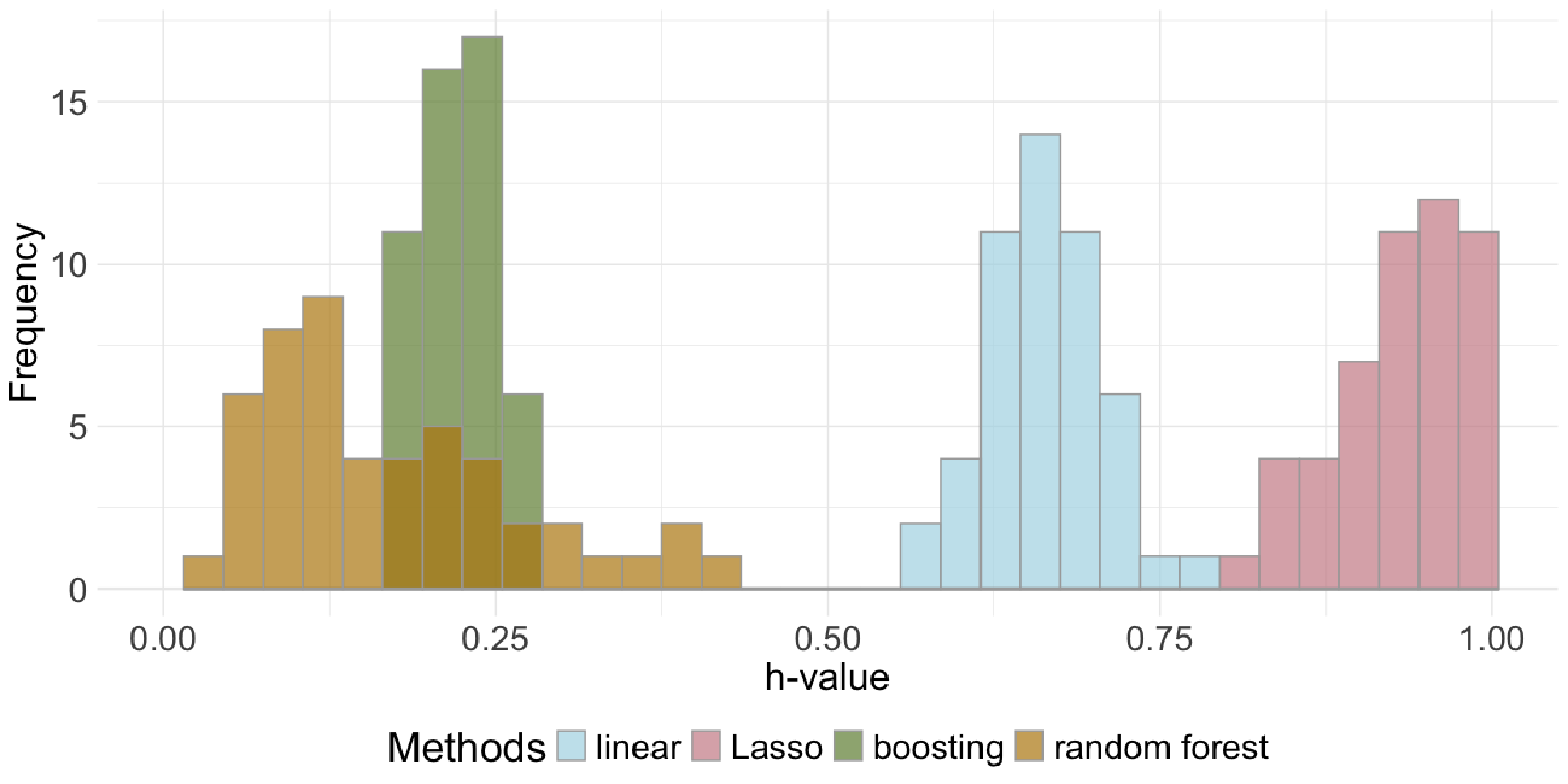}
    \caption{Histogram of h-values computed using linear regression, Lasso, boosting, and random forest approaches for the IHDP dataset. For each method, 50 h-values were computed. To compute each h-value, five-fold nested cross-validation and 50 repetitions were used.}
    \label{fig:histogram-hvalues-RF}
\end{figure}

Additional simulation results involving the random forest method are shown in Table \ref{tab:sim2-rf}
and Figure \ref{fig:histogram-hvalues-RF}.

\begin{table}[ht]
    \centering
    \begin{tabular}{ll|lccc}
   \multicolumn{2}{c}{} & \multicolumn{4}{c}{Method = random forest}\\
   \midrule
 $\mu(\cdot)$ & $\theta(\cdot)$ &$\theta_{XY}$ & Coverage & CI & Median\\
 &  & &   Proportion &Width &  One-sided h-value \\
 \midrule
 $\mu_1$ & $\theta_1$&-0.845&0.978&25.626&0.413\\
 $\mu_1$ & $\theta_2$&-0.492&0.996&25.522&0.426\\
 $\mu_1$ & $\theta_3$&-1.656&0.992&26.243&0.513\\
 $\mu_1$ & $\theta_4$&-5.351&0.944&82.215&0.347\\
 $\mu_2$ & $\theta_1$&-0.010&0.994&0.157&0.855\\
 $\mu_2$ & $\theta_2$&-0.107&0.988&0.218&0.958\\
 $\mu_2$ & $\theta_3$&-0.238&0.756&0.282&0.981\\
 $\mu_2$ & $\theta_4$&-0.274&0.832&16.822&0.114\\
 $\mu_3$ & $\theta_1$&-0.004&0.954&0.158&0.893\\
 $\mu_3$ & $\theta_2$&-0.108&0.990&0.229&0.952\\
 $\mu_3$ & $\theta_3$&-0.224&0.788&0.331&0.931\\
 $\mu_3$ & $\theta_4$&-0.371&0.886&17.579&0.129\\
 \bottomrule
    \end{tabular}
    \caption{Simulation results for random forest using the settings of the non-linear conditional mean function. All twelve combinations of the three baseline risk functions and four ITE functions described in Section 4.2 are considered. A sample size of $n = 500$ was used. The average value of the predictive estimand $\theta_{XY}$ across simulation replications is reported, and the coverage
and average width of the nested cross-validation based confidence intervals for $\theta_{XY}$ are shown for each setting. 
Median one-sided h-values across simulation replications are displayed for each simulation setting. }
    \label{tab:sim2-rf}
\end{table}

\section{Dataset Source}
The raw IHDP dataset can be downloaded from \\ \url{http://www.icpsr.umich.edu/icpsrweb/HMCA/studies/9795?paging.startRow=51}. 

\noindent
The dataset is titled: ``DS141: Transport Format SAS Library Containing the 59 Evaluation". The cleaned dataset containing the variables used in this study is available at \url{https://github.com/mahsaashouri/HTE-Model-Comparison}.

\end{document}